\documentclass[lettersize,journal]{IEEEtran}

\usepackage{amsmath,amsfonts}
\usepackage{algorithmic}
\usepackage{algorithm}
\usepackage{array}
\usepackage[caption=false,font=normalsize,labelfont=sf,textfont=sf]{subfig}
\usepackage{textcomp}
\usepackage{stfloats}
\usepackage{url}
\usepackage{verbatim}
\usepackage{graphicx}
\usepackage{cite}
\begin{document}

\title{Wall-Proximity Matters: Understanding the Effect of Device Placement with Respect to the Wall for Indoor Wi-Fi Sensing}

\author{He Wang,~\IEEEmembership{Student Member,~IEEE,}
        Yunpeng Ge,~\IEEEmembership{Student Member,~IEEE,}\\
        and Ivan Wang-Hei Ho,~\IEEEmembership{Senior Member,~IEEE}
\thanks{This work was supported in part by the Smart Traffic Fund (Project No. PSRI/31/2202/PR) established under the Transport Department of the Hong Kong Special Administrative Region (HKSAR), China.}
\thanks{The authors are with the Department of Electrical and Electronic Engineering, The Hong Kong Polytechnic University, Hong Kong, SAR, China (e-mail: edana.wang@connect.polyu.hk; yunpeng.ge@connect.polyu.hk; ivanwh.ho@polyu.edu.hk).}
\thanks{This work has been accepted for publication in the IEEE Internet of Things Journal. © 2025 IEEE. The published version is available at https://doi.org/10.1109/JIOT.2025.3637905. This is the author’s version of the manuscript submitted to IEEE.}
}

\maketitle

\pagestyle{empty}
\thispagestyle{empty}

\begin{abstract}

Wi-Fi sensing has been extensively explored for various applications, including vital sign monitoring, human activity recognition, indoor localization, and tracking. However, practical implementation in real-world scenarios is hindered by unstable sensing performance and limited knowledge of wireless sensing coverage. While previous works have aimed to address these challenges, they have overlooked the impact of walls on dynamic sensing capabilities in indoor environments. To fill this gap, we present a theoretical model that accounts for the effect of wall-device distance on sensing coverage. By incorporating both the wall-reflected path and the line-of-sight (LoS) path for dynamic signals, we develop a comprehensive sensing coverage model tailored for indoor environments. This model demonstrates that strategically deploying the transmitter and receiver in proximity to the wall within a specific range can significantly expand sensing coverage.  We assess the performance of our model through experiments in respiratory monitoring and stationary crowd counting applications, showcasing a notable 11.2\% improvement in counting accuracy. These findings pave the way for optimized deployment strategies in Wi-Fi sensing, facilitating more effective and accurate sensing solutions across various applications.  

\end{abstract}


\section{Introduction}

The recent exploration of channel state information (CSI) in Wi-Fi systems has revealed the untapped potential of communication signals for diverse indoor sensing applications \cite{rssi_csi_2013,survey2019}. In indoor environments, Wi-Fi signals encounter reflections from a multitude of objects, including walls and human bodies, resulting in complex multi-path propagation. Variations in CSI values predominantly arise from the dynamic movements of humans. This enables a wide range of sensing applications, including human activity recognition (HAR) \cite{humanactivity2023_2,HAR_TCCN_2024}, respiration monitoring \cite{vitalsign2024,vitalsign2025}, crowd counting \cite{crowdcounting2024,crowdcounting2025}, and indoor positioning \cite{indoorpositioning2021,indoorpositioning2023}. While prior work has made substantial progress in enhancing CSI-based sensing performance through advanced signal processing and deep learning techniques \cite{ratio2022,survey2023}, comparatively little effort has focused on improving the quality of the raw measurements. In practice, the effectiveness of sensing is often limited not by post-processing but by suboptimal deployment strategies, particularly the placement of devices, which directly shapes the strength and structure of multipath signals.

One commonly overlooked aspect of deployment is the placement of devices relative to surrounding walls, an element that can significantly influence multipath propagation, but has received limited attention in existing sensing models. In practical deployments, Wi-Fi transmitters, such as routers, are typically positioned near walls to avoid obstruction and facilitate wiring. However, prior work has primarily focused on the distance between the transmitter (Tx) and receiver (Rx) as the dominant placement factor \cite{placement_matters}, while largely neglecting the impact of walls on dynamic signal paths in indoor environments. Although some studies have examined the positive role of walls in through-wall sensing scenarios \cite{wall_matters}, these models assume that at least one device is located outside the room, thereby omitting the influence of indoor wall reflections. This reveals a critical gap in current understanding: the lack of a comprehensive model that accounts for indoor wall-reflected signals under common deployment scenarios. Bridging this gap is essential for optimizing sensing coverage in real-world environments.

\begin{figure*}[t!]
  \centering
  \includegraphics[width=0.7\linewidth]{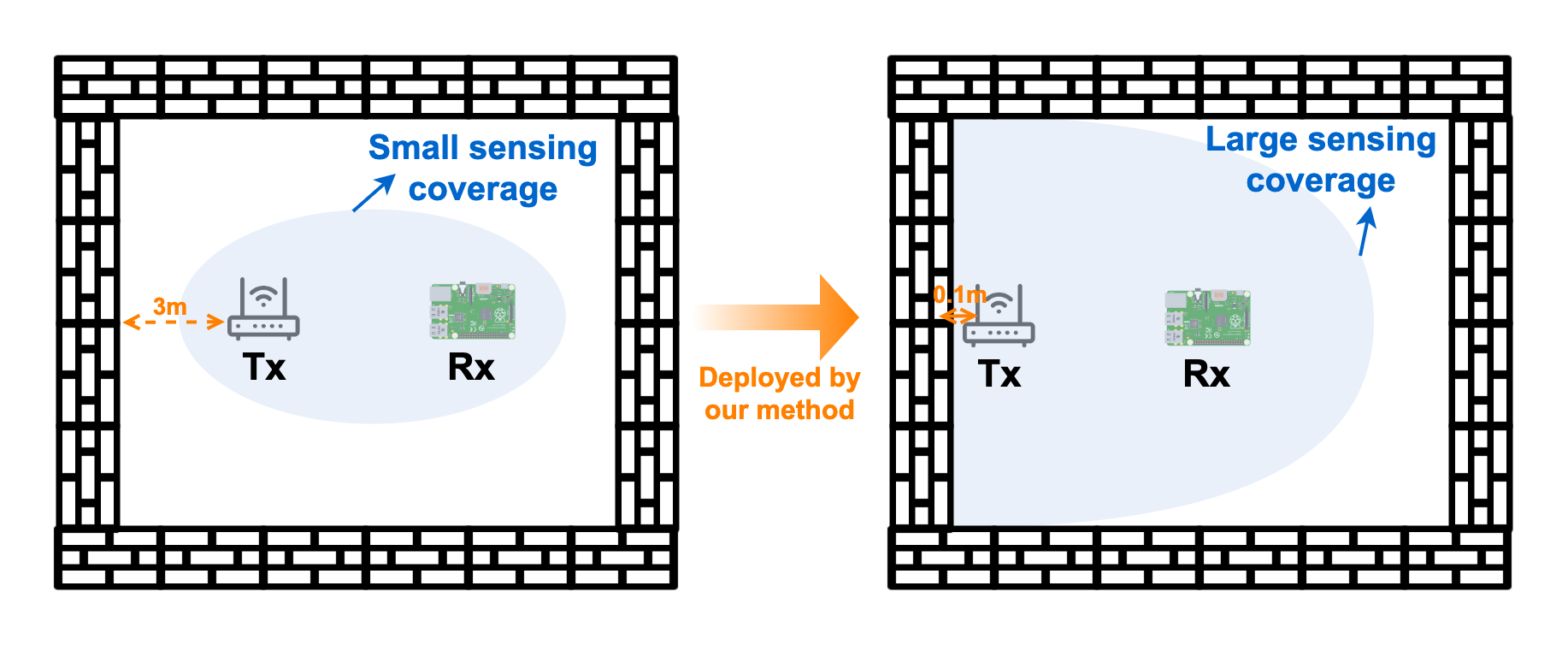}
  \caption{Enhancing sensing coverage through device deployment.}
  \label{intro_figure}
\end{figure*}

To address this gap, we propose the first analytical model that explicitly incorporates indoor wall-reflected dynamic signal paths into Wi-Fi sensing analysis. The model focuses on the dominant components, line-of-sight (LoS) and wall-reflected paths, to capture the primary interactions that shape sensing performance. While simplified by design, it remains lightweight, generalizable, and offers tractable insights into how wall proximity influences sensing coverage. This facilitates deployment-aware system design without relying on exhaustive simulation or dense empirical datasets. We complement the theoretical model with extensive experimental investigations to understand how wall placement affects real-world sensing performance. Results show that the distance between the wall and the sensing devices significantly impacts sensing coverage. As illustrated in Fig. \ref{intro_figure}, reducing the wall-to-transmitter distance can notably expand the effective sensing area. Interestingly, while wall reflections can enhance sensing coverage within the room, they may also increase unwanted sensitivity to movements beyond the wall, depending on transceiver placement. These findings highlight the potential to strategically leverage wall proximity to address key limitations in sensing coverage and robustness. This work represents a substantial step toward practical, deployment-optimized Wi-Fi sensing in real-world indoor environments.

The main contributions of this work are as follows:
\begin{itemize}
    \item This paper investigates wall reflections on dynamic signals in Wi-Fi systems within indoor environments, an area not previously explored. Our findings reveal a significant impact of walls on sensing, highlighting their potential to enhance wireless sensing capabilities. Given the presence of walls, our model represents a key advancement in developing device deployment strategies to optimize performance in real-world scenarios.

    \item We present a novel wall-reflection sensing model that incorporates both direct and reflected dynamic paths in indoor environments. Unlike prior models that only consider the direct path from devices to targets, our model's inclusion of wall-reflected paths expands the sensing region as the wall-device distance decreases, offering a more realistic representation of real-world scenarios. 
    
    \item This paper presents experimental results showcasing the practical effectiveness of our proposed model in expanding sensing coverage for respiratory monitoring and stationary crowd counting applications. By strategically positioning the devices with respect to the walls, we demonstrate a notable improvement of 11.2\% in the accuracy of stationary crowd counting. These results confirm the practical value of our model in guiding device placement and improving sensing performance. 

\end{itemize}

The rest of this paper is organized as follows. In Section II, we provide a brief overview of the related work in the field. Section III presents the preliminary knowledge related to the proposed model. The model itself, along with its properties, is detailed in Sections IV and V. In Section VI, we present the experimental results that validate the accuracy and usefulness of the proposed model. We discuss the limitations of our approach and potential directions for future work in Section VII, and conclude our work in Section VIII.

\section{Related Work}

This section provides a concise overview of existing research on the sensing coverage of RF-based contactless sensing systems, summarizing the key findings and developments in this field.

\subsection{Modeling sensing coverage in free space}

Extensive research has been conducted on the sensing coverage of radar systems, primarily focusing on free space conditions. Yang et al. \cite{radar2013} explored the geometrical relationship between the coverage area of a bistatic radar and the distance between the transmitter and receiver. Similar investigations have been carried out in Wi-Fi systems, where Xin et al. \cite{freesense2018} developed a sensing coverage model for different movement patterns in free space, utilizing a Fresnel zone model. The size of the sensing coverage was found to be influenced by the extent of the target's reflection surface, with larger reflection surfaces resulting in expanded sensing coverage. Wang et al. \cite{placement_matters} proposed a sensing coverage model based on sensing capability metrics, observing that coverage size initially increases but eventually decreases with increasing distances between the transmitter and receiver. 

While existing studies have delved into sensing coverage in indoor scenarios, none have considered the reflection effects of walls on dynamic signals. Typically, walls are viewed as obstacles in indoor environments that primarily affect static power within the sensing coverage model. However, the dynamic information from human activities is embedded in the wall-reflected signals. Hence, incorporating these reflected signals in the model can lead to a positive effect on sensing performance. This paper addresses the impact of walls on dynamic signals and explores the phase interactions between the wall-reflected path and the LoS path within the sensing coverage model.

\subsection{Modeling sensing coverage with blocks}

There have been research endeavors that consider non-line-of-sight (NLoS) scenarios, such as through-wall sensing, when developing sensing coverage models. Zhang et al. \cite{through_wall2023} proposed a through-wall wireless sensing model utilizing the Fresnel zone concept to characterize the sensing mechanisms of Wi-Fi signals in through-wall scenarios. Their refraction-aware Fresnel model demonstrated that the presence of walls can enhance through-wall sensing capabilities. Woodford et al. \cite{radar2022} employed curved reflectors and planar surfaces to achieve comprehensive coverage and enable NLoS radar sensing in crucial areas near intersections. Xie et al. \cite{wall_matters} developed a through-wall sensing model with LoRa systems, leveraging walls to expand sensing coverage and mitigate interference. 

The aforementioned studies focus on the impact of walls on through-wall sensing scenarios, where their models necessitate that at least one of the devices or targets be located on the other side of the wall, considering only a single signal path through the wall. In contrast to these works, our research delves into wall-reflected signals with both devices and targets positioned within the indoor environment, rather than the through-wall scenario. Our objective is to investigate how the integration of wall-reflected and LoS signals influences dynamic power, and thereby affects the sensing coverage region.

\section{Preliminaries}
In this section, we introduce the basics of CSI-based sensing, and the sensing coverage model without walls.

\subsection{Channel State Information}

\begin{figure}[h]
  \centering
  \includegraphics[width=1\linewidth]{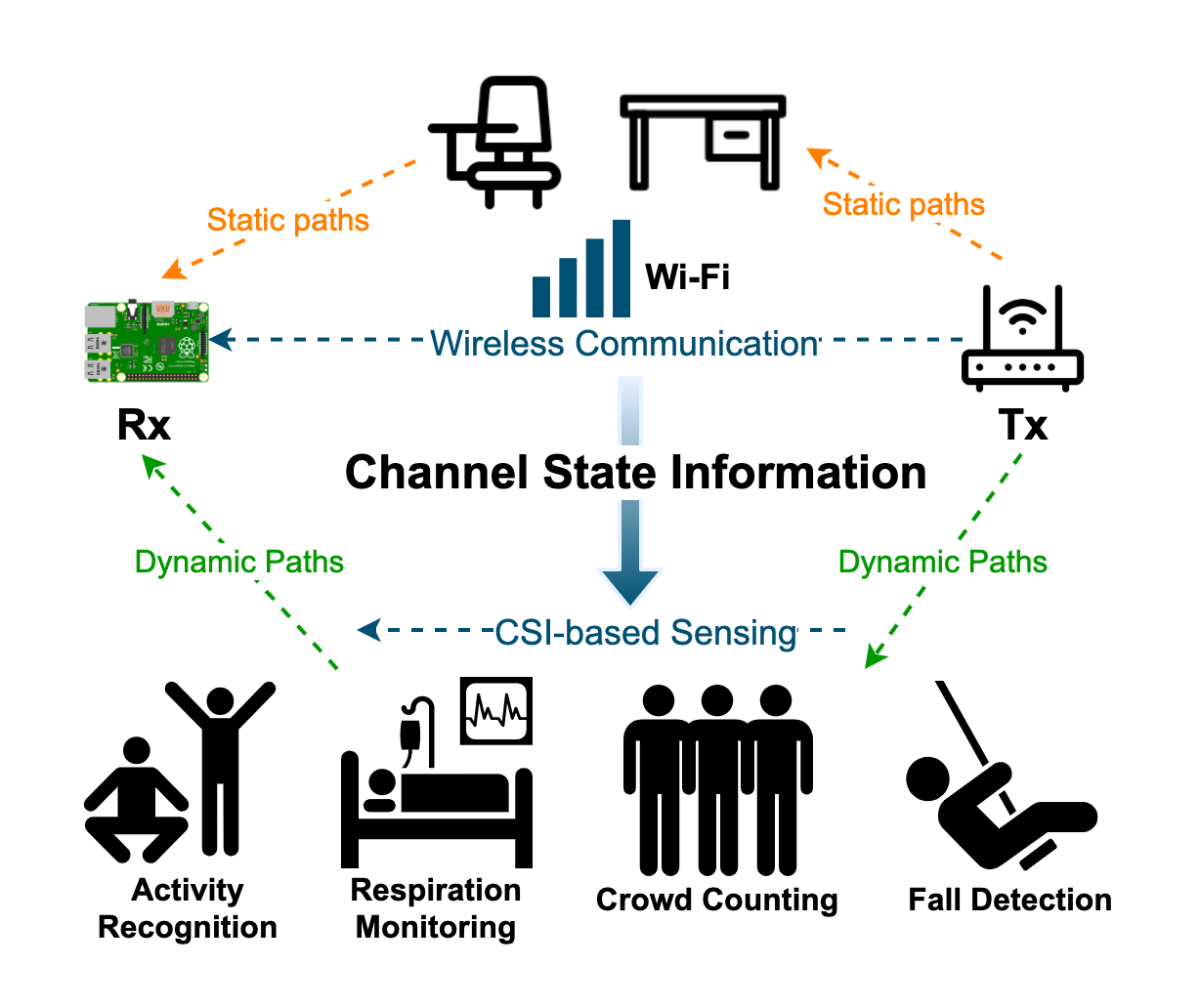}
  \caption{Categorizing Wi-Fi signal propagation: static paths vs dynamic paths.}
  \label{CSI_sensing}
\end{figure}

In an indoor environment, Wi-Fi signals can encounter numerous obstacles such as walls and human bodies, resulting in signal reflections and multiple paths to reach the receiver. These paths can be categorized into two types: static paths and dynamic paths, as depicted in Fig. \ref{CSI_sensing}. Static paths encompass the LoS path and paths reflected from stationary objects, while dynamic paths are generated by changes in target objects, such as human movements. CSI is utilized to characterize the wireless channel between a transmitter and a receiver in Wi-Fi systems \cite{fund2005}. The received signal can be mathematically represented by

\begin{equation}
    Y(f,t) = H(f,t) \times X(f,t), 
    \label{eq_csi1}
\end{equation}
where $X(f,t)$ and $H(f,t)$ denote respectively the frequency domain representations of the transmitted signal and the complex-valued channel frequency response (CFR) at carrier frequency $f$ and time $t$. Within the CSI matrix, each individual element can be represented as a linear combination of all the signal paths, which can be demonstrated as

\begin{equation}
\begin{aligned}
    H(f,t) &= H_{s}(f,t) + H_{d}(f,t) + H_{n}(f,t) \\
    &= \left | H_{s}(f,t) \right |e^{-j\theta_{s}} + \left | H_{d}(f,t) \right |e^{-j\theta_{d}} \\
    &+ \left | H_{n}(f,t) \right |e^{-j\theta_{n}}, 
\end{aligned}
\label{eq_csi2}
\end{equation}
where $H_{s}$ denotes the contribution of the static path, $H_{d}$ represents the dynamic path, and $H_{n}$ corresponds to the noise component. The amplitude of each CSI component is denoted by $|H|$, while $\theta$ indicates the phase \cite{csimodel2015}. By decomposing the CSI matrix in this manner, we can effectively analyze and understand the contributions of different signal paths to the overall wireless channel characteristics.

\subsection{Sensing Coverage Model without Walls}
\label{section_nowall}

A metric named SSNR (sensing-signal-to-noise-ratio) was proposed by Wang et al. \cite{placement_matters} to quantify the sensing capability:

\begin{equation}
    SSNR = \frac{P_{d} }{P_{i} } =\frac{P_{d} }{\gamma P_{s}+b }, 
\label{SSNR_define}
\end{equation}
where $P_{d}$ is the power of the dynamic signal, $P_{i}$ is the interference power, which is linearly proportional to the static power $P_{s}$. According to the findings in \cite{placement_matters}, the variables $\gamma$ and $b$ have fixed values for a given pair of Wi-Fi transceivers. It is worth noting that the static path is assumed to be the LoS path, and the power of the LoS signal can be mathematically expressed using the Friis equation \cite{friis1946, friis2018, friis2020} as

\begin{equation}
     P_{s }=\frac{P_{T}G_{T}A_{R}}{4\pi (r_{D} )^{2} },
\label{power_los}
\end{equation}
where $P_{T}$ represents the transmission power, $G_{T}$ indicates the antenna gain of the transmitter, and $r_{D}$ represents the distance between the transmitter and receiver. $A_{R}$ corresponds to the effective aperture of the receiver's antenna, which can be calculated as $A_{R}=\frac{G_{R}\lambda^{2}}{4\pi}$, with $G_{R}$ denoting the antenna gain of the receiver and $\lambda$ representing the wavelength of the signal. However, in real indoor scenarios, the static path includes reflections from other stationary objects, which can lead to variations in the interference power $P_{i}$ \cite{placement_matters}. 

\begin{figure}[h]
  \centering
  \includegraphics[width=0.7\linewidth]{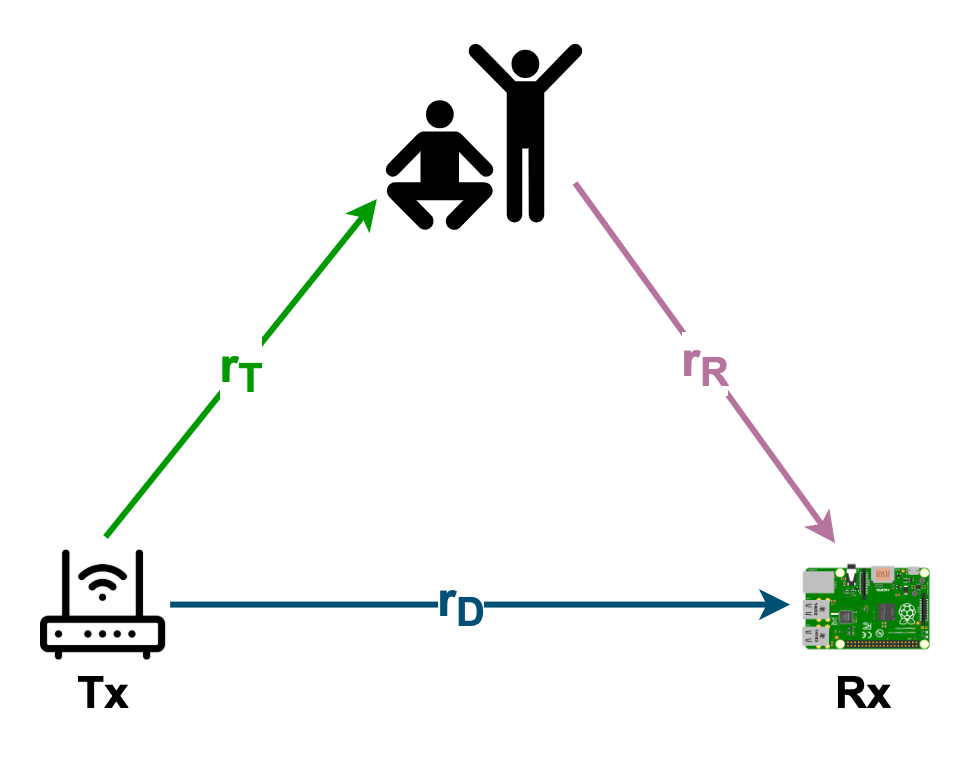}
  \caption{Sensing model without walls.}
  \label{paths_noWalls}
\end{figure}

Consider a simple scenario in free space, the dynamic path can be divided into two parts as shown in Fig. \ref{paths_noWalls}. The first part with a distance $r_{T}$ is the signal from the transmitter to the target, and the second part with a distance $r_{R}$ is the reflected signal from the target to the receivers. According to the Friis equation \cite{friis1946}, the power of the signal arriving at the target can be expressed as

\begin{equation}
    P_{r_{T} }=\frac{P_{T}G_{T}}{4\pi (r_{T} )^{2} }.    
\label{power_nowall_1}
\end{equation}

The power of the signal reflected by the target and received by the receiver can be mathematically expressed as

\begin{equation}
    P_{d_{LoS} }=\frac{P_{r_{T}}\sigma A_{R}}{4\pi (r_{R} )^{2} }=\frac{P_{T}G_{T}\sigma A_{R}}{(4\pi)^{2} (r_{T}r_{R} )^{2}},        
\label{power_nowall_2}
\end{equation}
where $\sigma$ represents the radar cross section (RCS) of the target, encompassing the effective reflection ratio at the target \cite{RCS2015}. In free space, the power of the dynamic signal $P_{d}$ equals to $P_{d_{LoS}}$. Therefore, the $SSNR_{LoS}$ can be expressed as

\begin{equation}
    SSNR_{LoS} =  \frac{P_{d_{LoS}} }{\gamma P_{s}+b }
 =\frac{K\sigma }{(4\pi) (r_{T}r_{R} )^{2}(\gamma \frac{K}{r_{D}^{2}} +b)},  
\label{SSNR_nowall}
\end{equation}
where $K=\frac{P_{T}G_{T}A_{R}}{4\pi }$. Assuming a constant RCS of the target $\sigma$ and a small value for $b$. Furthermore, the transmission power $P_{T}$, antenna gains $G_{T}$ and $G_{R}$, and signal wavelength $\lambda$ are assumed to be constant, then \eqref{SSNR_nowall} can be simplified as

\begin{equation}
    SSNR_{LoS} \propto  \frac{r_{D}^2}{(r_{T}r_{R})^2}. 
\label{SSNR_nowall_simple}
\end{equation}

The derived equation demonstrates the close relationship between the sensing capability and various distances, namely the distances of the target reflection signal ($r_{T}$ and $r_{R}$) and the distance between the transceivers ($r_{D}$). By considering the minimum SSNR requirement, we can express the boundary of the sensing area as

\begin{equation}
    (r_{T}r_{R})_{b} \propto  \sqrt{\frac{r_{D}^2}{SSNR_{min}} }.  
\label{SSNR_nowall_bound}
\end{equation}

Note that the minimum SSNR requirement $SSNR_{min}$ varies with sensing applications and signal preprocessing methods. From \eqref{SSNR_nowall_bound}, we can see that the sensing area initially increases and then decreases as the length of the LoS path increases.

\section{Understanding the effect of walls on indoor sensing systems}
\label{Section_IV}

The sensing coverage model outlined in \cite{placement_matters} primarily concentrates on free space scenarios, acknowledging walls as obstacles that solely impact static power. Nevertheless, when devices or targets are close to walls, the influence of signals reflected by walls on dynamic power becomes significant. In response, we have developed a novel model customized for multipath-dominant indoor environments. This model accounts for the impact of walls and factors in the distances between the Tx and Rx, to precisely evaluate indoor sensing capabilities.

\begin{figure*}[t!]
\centering
\subfloat[]{\includegraphics[width=2.1in]{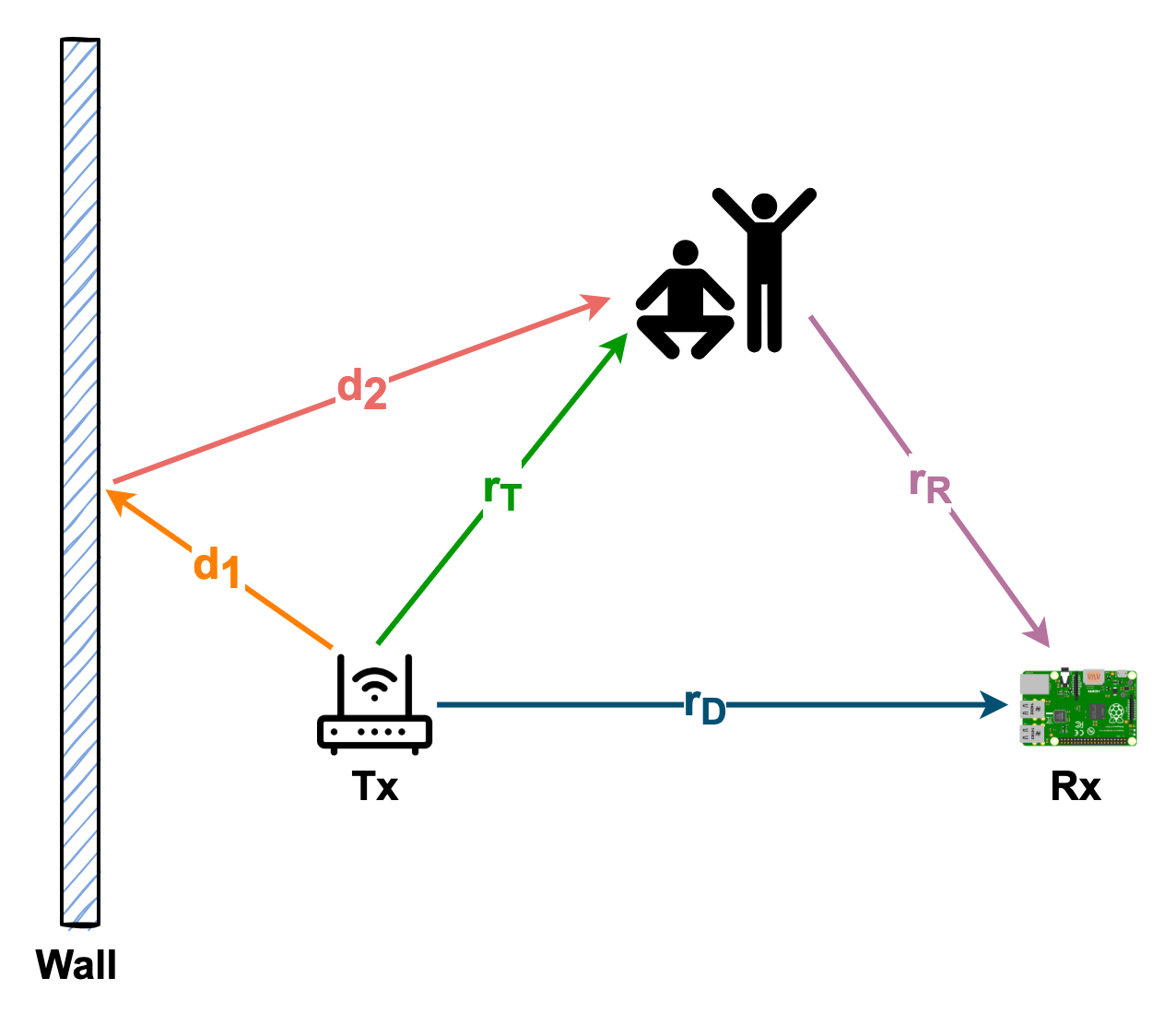}%
\label{wall_left}}
\subfloat[]{\includegraphics[width=2.1in]{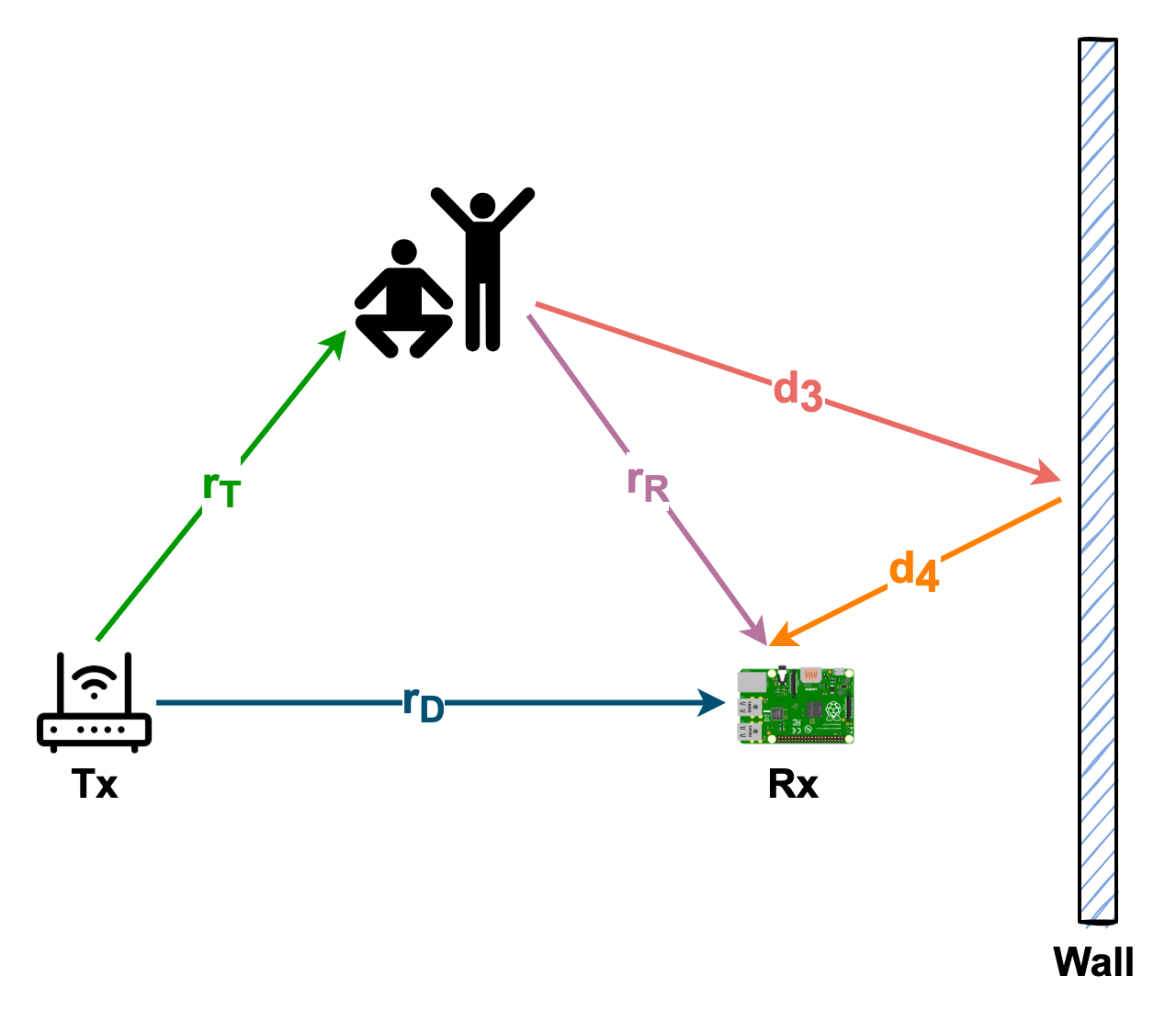}%
\label{wall_right}}
\subfloat[]{\includegraphics[width=2.58in]{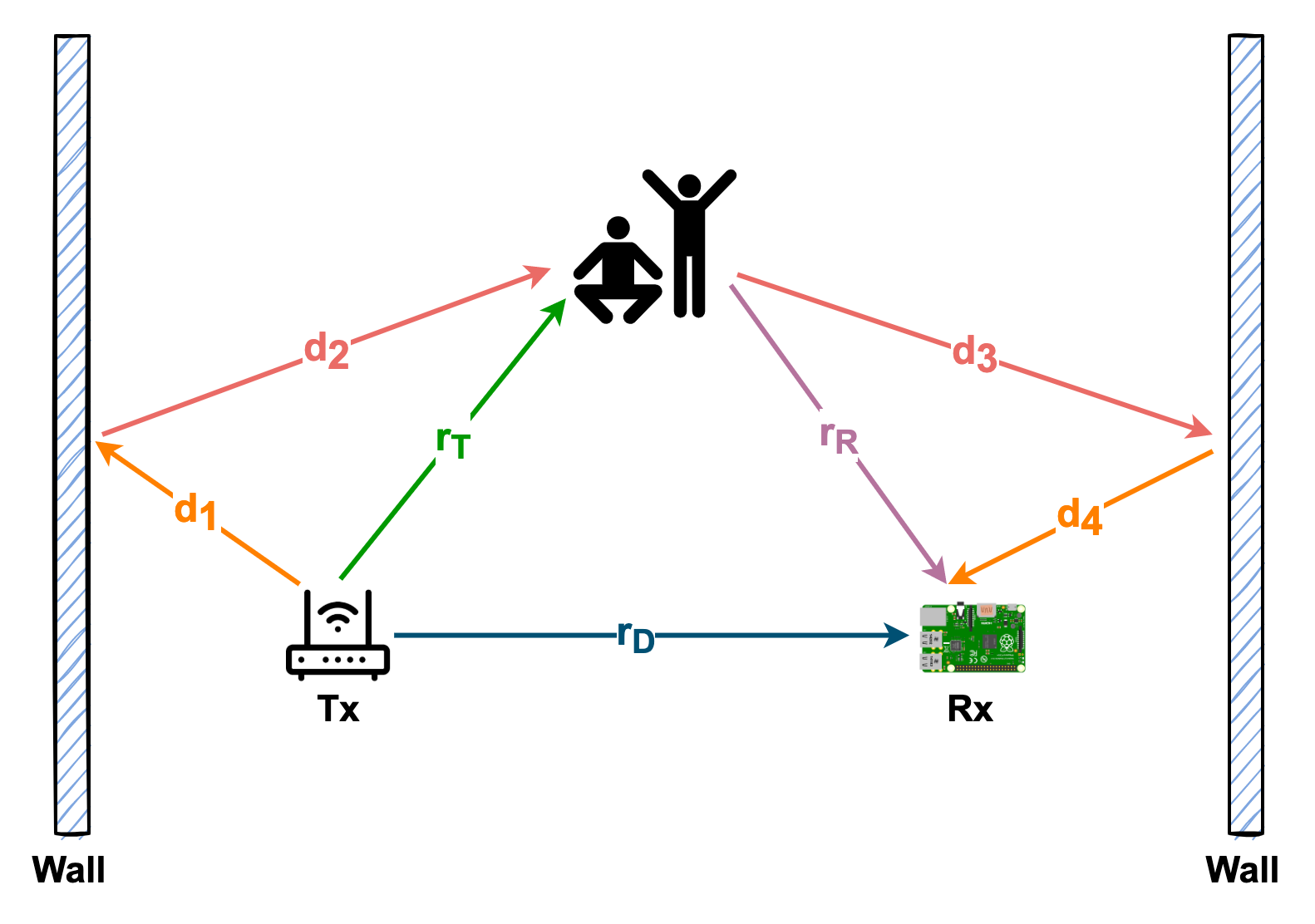}%
\label{2walls}}
\caption{Wall-reflection sensing model. (a) Tx near the wall. (b) Rx near the wall. (c) Both Tx and Rx near walls.}
\label{paths_Wall_reflection}
\end{figure*}

Fig. \ref{paths_Wall_reflection} illustrates three common wall-reflection scenarios encountered in indoor environments: i) when the transmitter is deployed in close proximity to the wall (Fig. \ref{wall_left}); ii) when the receiver is deployed in close proximity to the wall (Fig. \ref{wall_right}); and iii) when both the transmitter and receiver are deployed in close proximity to the wall (Fig. \ref{2walls}). In practical scenarios, receivers are often mobile sensing devices such as smartphones and tablets, which are typically user-controlled and rarely deployed near walls. In contrast, routers, functioning as transmitters, are typically deployed in a controlled manner and are commonly placed close to walls. Therefore, this paper primarily focuses on the scenario depicted in Fig. \ref{wall_left} to demonstrate the impact of wall reflection. However, the theories and findings presented in this paper are applicable to the scenarios in Fig. \ref{wall_right} and Fig. \ref{2walls} as well. 

\subsection{Wall-reflection Sensing Model}

When there is a wall close to the transmitter, parts of the signals will be reflected by the wall before arriving at the target as shown in Fig. \ref{wall_left}. We can divide the wall-reflected signal into two parts: from the Tx to the wall with a distance $d_{1}$ and reflected by the wall to the target with a distance $d_{2}$. Based on the Friis equation \cite{friis1946,friis2018,friis2020}, the power of the signal arriving at the wall can be expressed as

\begin{equation}
    P_{d_{1} }=\frac{P_{T}G_{T}}{4\pi (d_{1} )^{2} }.    
\label{power_reflectwall_1}
\end{equation}

The power of the signal reflected by the wall to the target can be mathematically expressed as

\begin{equation}
    P_{d_{2} }=\frac{P_{d_{1}} R_{wall}^2}{4\pi (d_{2} )^{2} }=\frac{P_{T}G_{T}R_{wall}^2}{(4\pi)^{2} (d_{1}d_{2} )^{2}},        
\label{power_reflectwall_2}
\end{equation}
where $R_{wall}$ is the reflection coefficient of the wall \cite{reflection1948,reflection1996}. The power of the signal reflected by the target and arriving at the receiver can be expressed as 

\begin{equation}
    P_{d_{wall} }=\frac{P_{d_{2}}\sigma A_{R}}{4\pi (r_{R} )^{2} }=\frac{P_{T}G_{T}R_{wall}^2\sigma A_{R}}{(4\pi)^{3} (d_{1}d_{2}r_{R} )^{2}}.        
\label{power_reflectwall_3}
\end{equation}

Assume that the static path is the LoS path, and the power of the LoS signal is the same as \eqref{power_los}:

\begin{equation}
     P_{s }=\frac{P_{T}G_{T}A_{R}}{4\pi (r_{D} )^{2} }.
\label{power_reflectwall_los}
\end{equation}

To incorporate the impact of wall reflections on indoor environments, the $SSNR_{LoS}$ proposed in \cite{placement_matters} can be adapted using the following formula:

\begin{equation}
    SSNR_{wall} = \frac{P_{d_{wall}} }{\gamma P_{s}+b } =\frac{KR_{wall}^2\sigma }{(4\pi)^2 (d_{1}d_{2}r_{R} )^{2}(\gamma \frac{K}{r_{D}^{2}} +b)}. 
\label{SSNR_reflectwall}
\end{equation}

By considering that $K$ and $\gamma$ are constants, and assuming a constant RCS of the target $\sigma$, a constant reflection coefficient of the wall $R_{wall}$, and a small value for $b$, \eqref{SSNR_reflectwall} can be simplified as

\begin{equation}
    SSNR_{wall} \propto  \frac{r_{D}^2}{(d_{1}d_{2}r_{R} )^2}. 
\label{SSNR_reflectwall_simple}
\end{equation}

From this equation, we can see that the sensing capability is closely related to the distances of the target-reflected signal ($d_{1}$, $d_{2}$, and $r_{R}$) and the distance of the static signal ($r_{D}$). It also provides a means to investigate how the sensing capability changes with varying target locations. Notably, regions close to the devices and the wall demonstrate higher sensing capabilities. This implies that targets situated in close proximity to these regions are more likely to be accurately detected.

However, it is important to note that this model solely considers the dynamic signals resulting from the reflection off the wall. In Section \ref{section_reflectwall}, we will further explore the impact of the dominant target reflected signals originating from the LoS signals.

\subsection{The Sensing Coverage Model in Indoor Environments Considering the Effect of Walls}
\label{section_reflectwall}

While the presence of walls does affect the dynamic power in indoor environments, it is important to note that the paths with $r_{T}$ and $r_{R}$ depicted in Fig. \ref{paths_Wall_reflection} continue to play a dominant role in the dynamic paths. To create a comprehensive model for indoor environments, we can incorporate the wall-reflected paths $d_{1}$ and $d_{2}$ along with the existing paths to calculate the overall dynamic power accurately. By considering these combined paths, we can develop a comprehensive model that accounts for both LoS and wall-reflected signals in indoor environments.

According to \cite{humanloc2016}, the combined dynamic power can be expressed as  

\begin{equation}
\begin{aligned}
P_{d}&=P_{d_{wall}}+P_{d_{LoS}}+2\sqrt{P_{d_{wall}} P_{d_{LoS}}}\cos (\bigtriangleup \phi)  \\
&=\frac{P_{T}G_{T}R_{wall}^2\sigma A_{R}}{(4\pi)^{3} (d_{1}d_{2}r_{R} )^{2}}+\frac{P_{T}G_{T}\sigma A_{R}}{(4\pi)^{2} (r_{T}r_{R} )^{2}}\\
&+\frac{2P_{T}G_{T}R_{wall}\sigma A_{R}\cos(\bigtriangleup \phi ) }{(4\pi)^{\frac{5}{2} }(r_{R})^{2} (d_{1}d_{2}r_{T} )},     
\label{power_comb}
\end{aligned}
\end{equation}
where $P_{d_{wall}}$ is expressed in \eqref{power_reflectwall_3} and $P_{d_{LoS}}$ is expressed in \eqref{power_nowall_2}. The phase difference between the two signals can be expressed as $\bigtriangleup \phi  = \frac{2\pi(d_{1}+d_{2}-r_{T} ) }{\lambda }$, where $\lambda$ is the signal wavelength. 

Eq. \eqref{power_comb} gives the combined dynamic power considering the LoS and the main-wall reflection only. However, in practical indoor environments, additional weak reflections may arise from side walls or other objects. To evaluate their influence, an error bound analysis is conducted to assess the effect of ignoring such reflections and to ensure that the simplified single-wall assumption does not alter the overall conclusions. Specifically, the additional dynamic power contributed by the side-wall reflection can be expressed as

\begin{equation}
\begin{aligned}
\Delta P &= P_{d_{side}} + 2 \sqrt{P_{d_{LoS}} P_{d_{side}}} 
\cos(\Delta \phi_{side}) \\
&+ 2 \sqrt{P_{d_{wall}} P_{d_{side}}}
\cos(\Delta \phi_{side} - \Delta \phi_{\text{wall}}),
\label{power_side_single}
\end{aligned}
\end{equation}
where $\Delta P$ denotes the side-wall contribution beyond the baseline model in \eqref{power_comb}. Here, $P_{d_{side}}$ follows the same expression as $P_{d_{wall}}$ in \eqref{power_reflectwall_3}, with distances $d_3$ and $d_4$ and reflection coefficient $R_{side}$, while $\Delta \phi_{side}$ is defined as
$\Delta \phi_{side} = \frac{2\pi \big((d_3+d_4)-r_T\big)}{\lambda}$. The total dynamic power including the side-wall reflection can then be expressed as
\begin{equation}
P_{full} = P_{d} + \Delta P.
\label{power_full}
\end{equation}

To quantify the approximation error introduced by ignoring the side-wall contribution, we define the relative error between the full and simplified models as
\begin{equation}
\mathrm{RelErr} = \frac{|P_{full} - P_{d}|}{P_{full}}.
\label{relerr}
\end{equation}

As a concrete example, we evaluate \eqref{relerr} in the meeting room used for data collection (room size $6.3\,\text{m}\times 8.4\,\text{m}$), where the transmitter is placed $1\,\text{m}$ from the main wall and $3.15\,\text{m}$ from the side wall. The average relative error refers to the spatial average of $\mathrm{RelErr}$ when the human target is placed at multiple positions across the room area. To obtain an upper bound, all paths are assumed to be in phase ($\cos(\Delta\phi)=1$). Under this configuration, the average relative error is $3.26\%$ (approximately $0.14\,\text{dB}$). These results indicate that, for this practical deployment, ignoring the side-wall contribution introduces only minor deviations.

Given that the side-wall contribution is shown to be negligible, we proceed with the simplified single-wall model in the subsequent SSNR analysis. As the static power $P_{s}$ is shown in \eqref{power_reflectwall_los}, the SSNR can be expressed as

\begin{equation}
\begin{aligned}
    SSNR &= \frac{P_{d} }{P_{i} } =  \frac{P_{d_{wall}}+P_{d_{LoS}}+2\sqrt{P_{d_{wall}} P_{d_{LoS}}}\cos (\bigtriangleup \phi) }{\gamma P_{s}+b } \\
    &=\frac{KR_{wall}^2\sigma }{(4\pi)^2 (d_{1}d_{2}r_{R} )^{2}(\gamma \frac{K}{r_{D}^{2}} +b)}\\
    &+\frac{K\sigma }{4\pi (r_{T}r_{R} )^{2}(\gamma \frac{K}{r_{D}^{2}} +b)}\\
    &+\frac{2KR_{wall}\sigma \cos(\bigtriangleup \phi ) }{(4\pi)^{\frac{3}{2} }(r_{R})^{2} (d_{1}d_{2}r_{T} )(\gamma \frac{K}{r_{D}^{2}} +b)}.       
\label{SSNR_comb}
\end{aligned}
\end{equation}

By considering that $K$ and $\gamma$ are constants, and assuming a constant RCS of the target $\sigma$, and a small value for $b$, \eqref{SSNR_comb} can be simplified as

\begin{equation}
\begin{aligned}
    SSNR &\propto \frac{\alpha_{1} (r_{D})^2 }{(d_{1}d_{2}r_{R} )^{2}}+\frac{(r_{D})^2 }{(r_{T}r_{R} )^{2}}+\frac{\alpha_{2}cos(\bigtriangleup \phi ) (r_{D})^2 }{(d_{1}d_{2}r_{T})(r_{R} )^{2}}\\
    &=\alpha_{1} SSNR_{wall}+SSNR_{LoS}\\
    &+\alpha_{2}\frac{cos(\bigtriangleup \phi ) (r_{D})^2 }{(d_{1}d_{2}r_{T})(r_{R} )^{2}}, 
\end{aligned}  
\label{SSNR_comb_simple}
\end{equation}
where $SSNR_{wall}$ is shown in \eqref{SSNR_reflectwall_simple} and $SSNR_{LoS}$ is shown in \eqref{SSNR_nowall_simple}. Parameters $\alpha_{1}=\frac{R_{wall}^2}{4\pi} $ and $\alpha_{2}=\frac{R_{wall}}{\sqrt{\pi}} $, where the reflection coefficient $R_{wall}$ varies with different wall materials and incident angles \cite{reflection1996}.

\begin{figure}[h]
  \centering
  \includegraphics[width=0.8\linewidth]{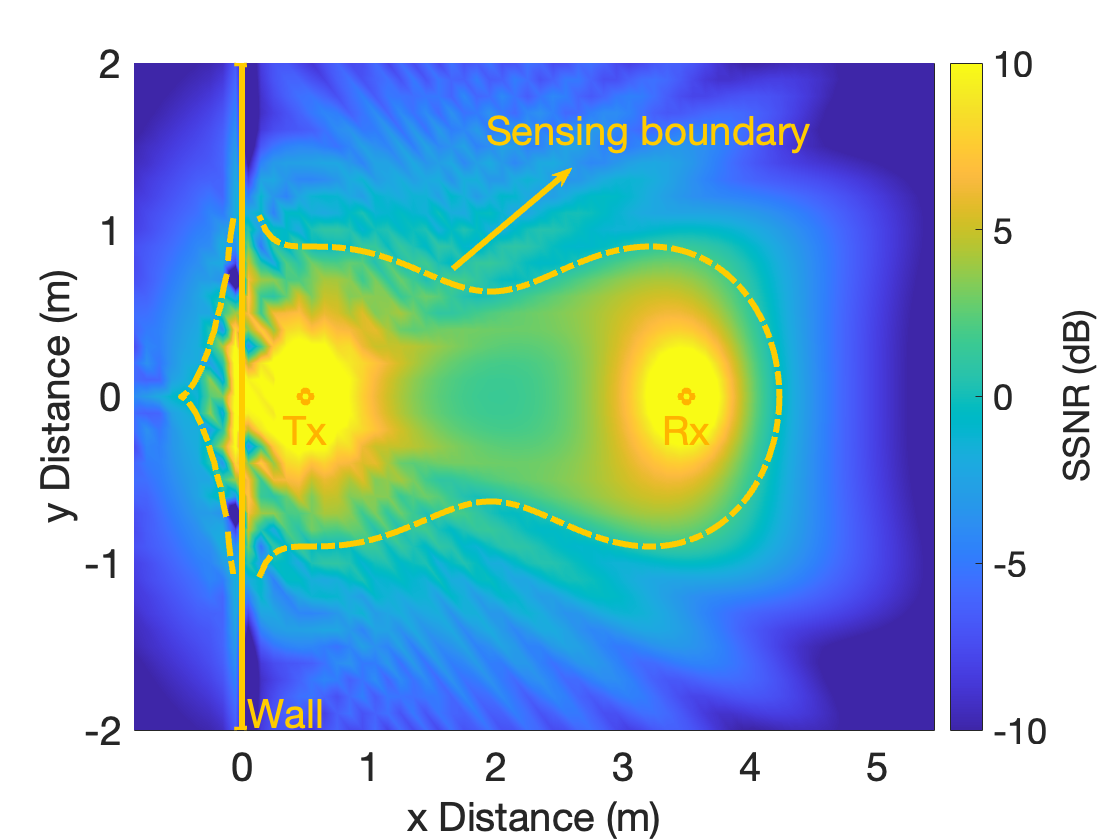}
  \caption{Heatmap of sensing capability in indoor environments.}
  \label{heatmap_comb}
\end{figure}

Fig. \ref{heatmap_comb} illustrates the heatmap of the sensing capability based on our proposed model in \eqref{SSNR_comb_simple}, with a wall–transmitter distance of 0.5 m. The model integrates dynamic signals from both LoS conditions and wall reflections. As analyzed, variations in the reflection coefficient have limited influence on the model characteristics; thus, we adopt a fixed reflection coefficient of 0.3 \cite{reflection_2017}. Since this work primarily investigates the impact of indoor wall reflections on sensing performance, we do not explicitly model through-wall propagation. Instead, following a simplified geometric modeling perspective inspired by \cite{wall_matters}, we use a common attenuation coefficient for both reflection and (if any) penetration paths (i.e., assuming $R_{\text{wall}} = T_{\text{wall}}$). This simplification does not affect our core conclusions, as the sensing capability analysis mainly focuses on regions inside or adjacent to the wall. In the heatmap, yellow indicates higher sensing capability, while blue represents lower capability. The rapid fluctuations between light and dark regions result from phase interactions between the two paths. However, this effect is minor compared to the dominant sensing regions near the wall and transceivers. Overall, the heatmap shows that regions close to devices and the wall exhibit a higher sensing capability, suggesting that targets in these areas are more likely to be detected.

For the sensing boundary, we consider a minimum SSNR requirement of 2 dB for reliable sensing, as established in previous works \cite{placement_matters, wall_matters}. Due to the phase interplay between the two signal paths, the actual sensing boundaries manifest as zigzag patterns. To facilitate easier computation of the sensing area, the visualized sensing boundary is smoothened, as demonstrated in Fig. \ref{heatmap_comb}. By examining \eqref{SSNR_comb_simple}, we can observe that both the Tx-Rx distance and the wall-device distance play significant roles in determining the sensing boundary. The area of the region with the same SSNR expands as the device gets closer to the wall. In Section \ref{Section_analysis}, we will delve deeper into the analysis of this sensing boundary, considering the aforementioned distances to provide additional insights.

\section{Enhancing Sensing Coverage in Indoor Environments with Walls}
\label{Section_analysis}

In this section, we investigate the fundamental factors that impact the sensing coverage in the presence of wall reflections by illustrating the theoretical sensing-coverage model derived in Section \ref{Section_IV} with numerical simulations. The analytical results presented here serve as theoretical validation before experimental verification in Section \ref{Section_VI}. Building on this theoretical analysis, we aim to understand how these factors can be effectively managed to address two prominent issues in wireless sensing: limited sensing coverage and instability due to interference. By analyzing and controlling the sensing coverage, we can mitigate the challenges posed by these issues and enhance the overall performance of wireless sensing systems.

\subsection{Factors affecting sensing coverage}
\label{subsection_factors}

According to the sensing coverage model discussed in Section \ref{section_reflectwall}, the shape and size of the sensing coverage are significantly influenced by two primary factors: the distance between the wall and the devices and the distance between the transmitter and receiver. 

{\bfseries Factor I: Wall-Device Distance.}
\label{Section_Wall_Tx}

\begin{figure*}[!t]
\centering
\subfloat[]{\includegraphics[width=2.1in]{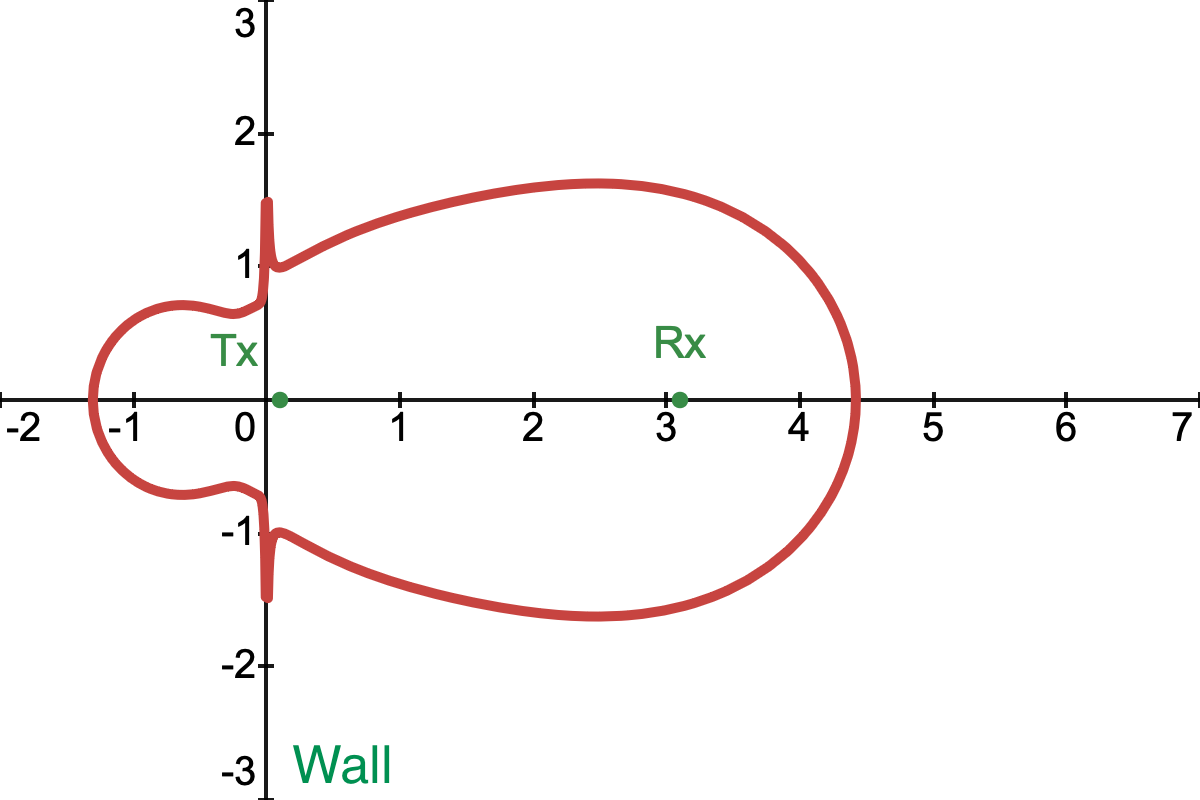}%
\label{wall_0.1}}
\subfloat[]{\includegraphics[width=2.1in]{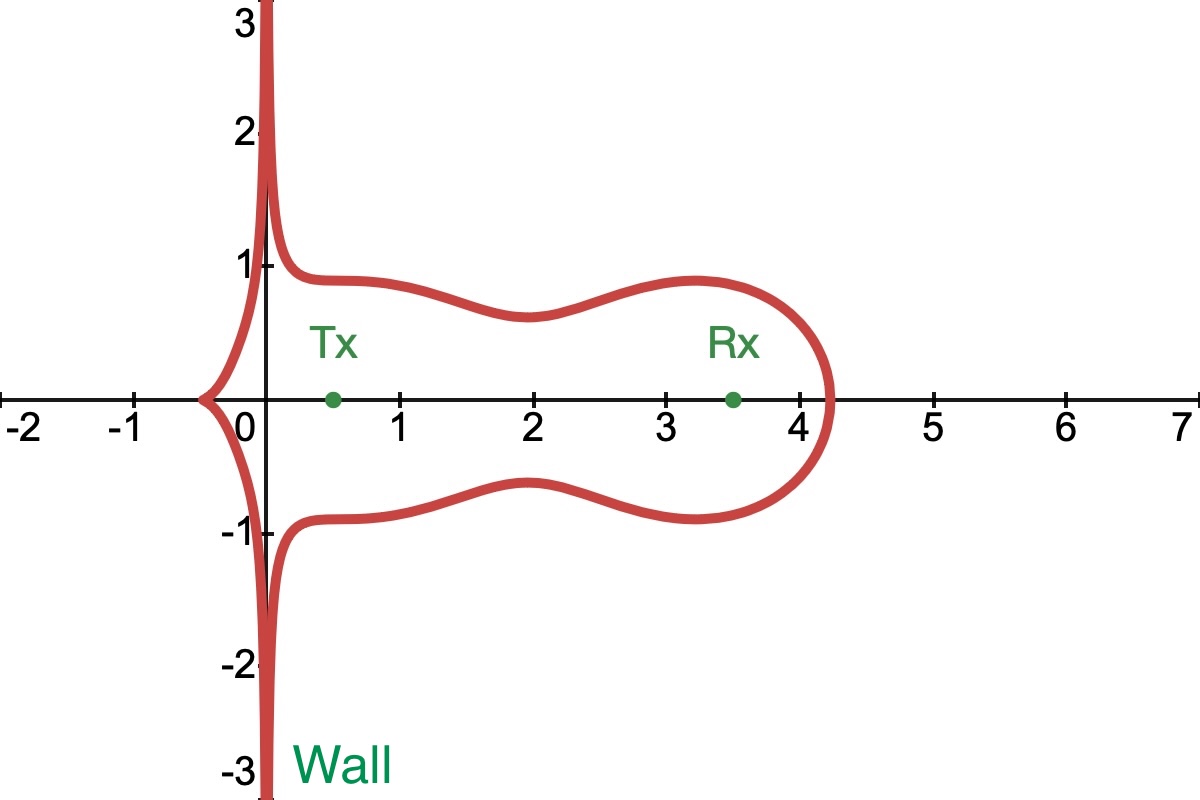}%
\label{wall_0.5}}
\subfloat[]{\includegraphics[width=2.1in]{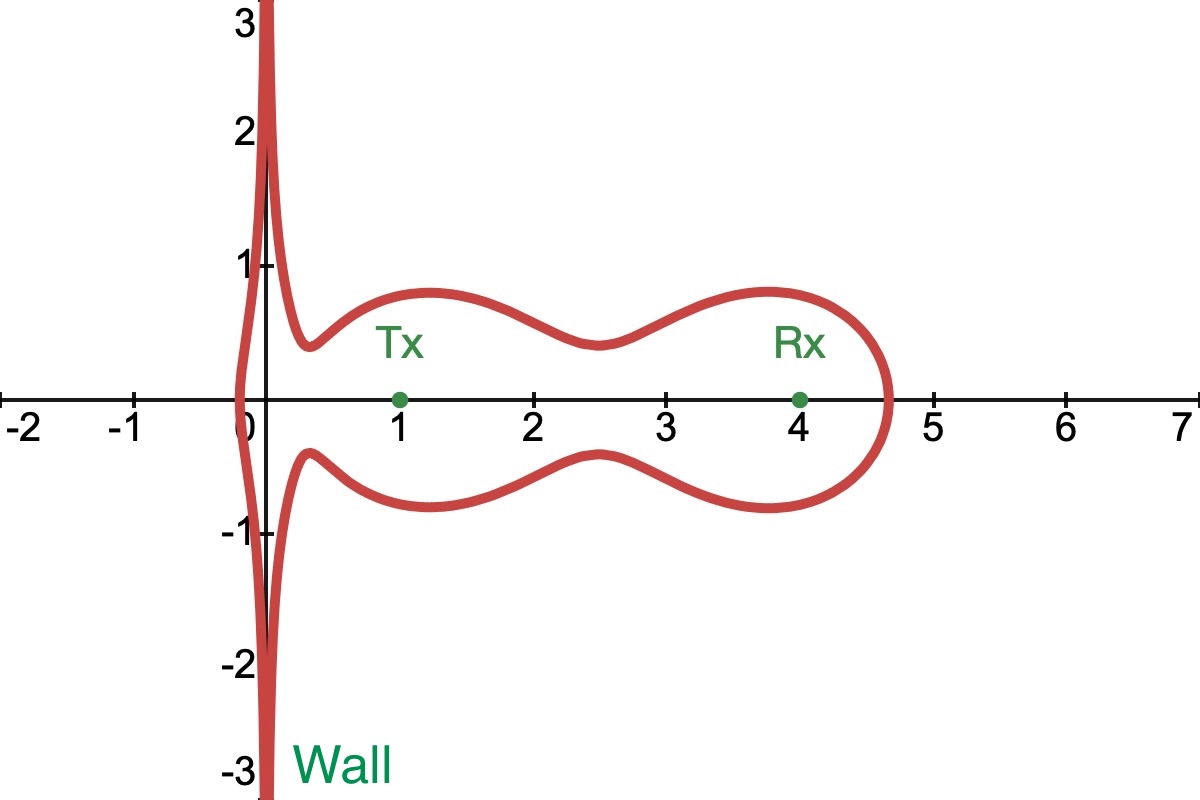}%
\label{wall_1}}

\subfloat[]{\includegraphics[width=2.1in]{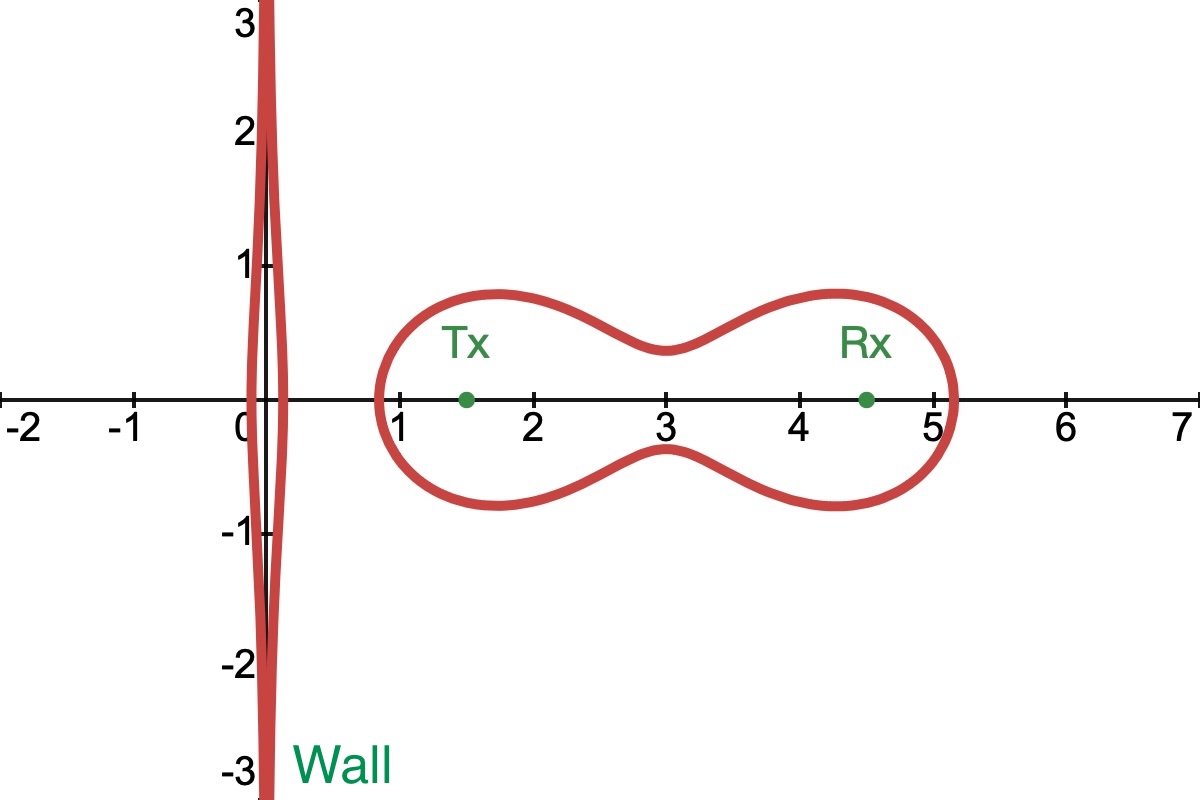}%
\label{wall_1.5}}
\subfloat[]{\includegraphics[width=2.1in]{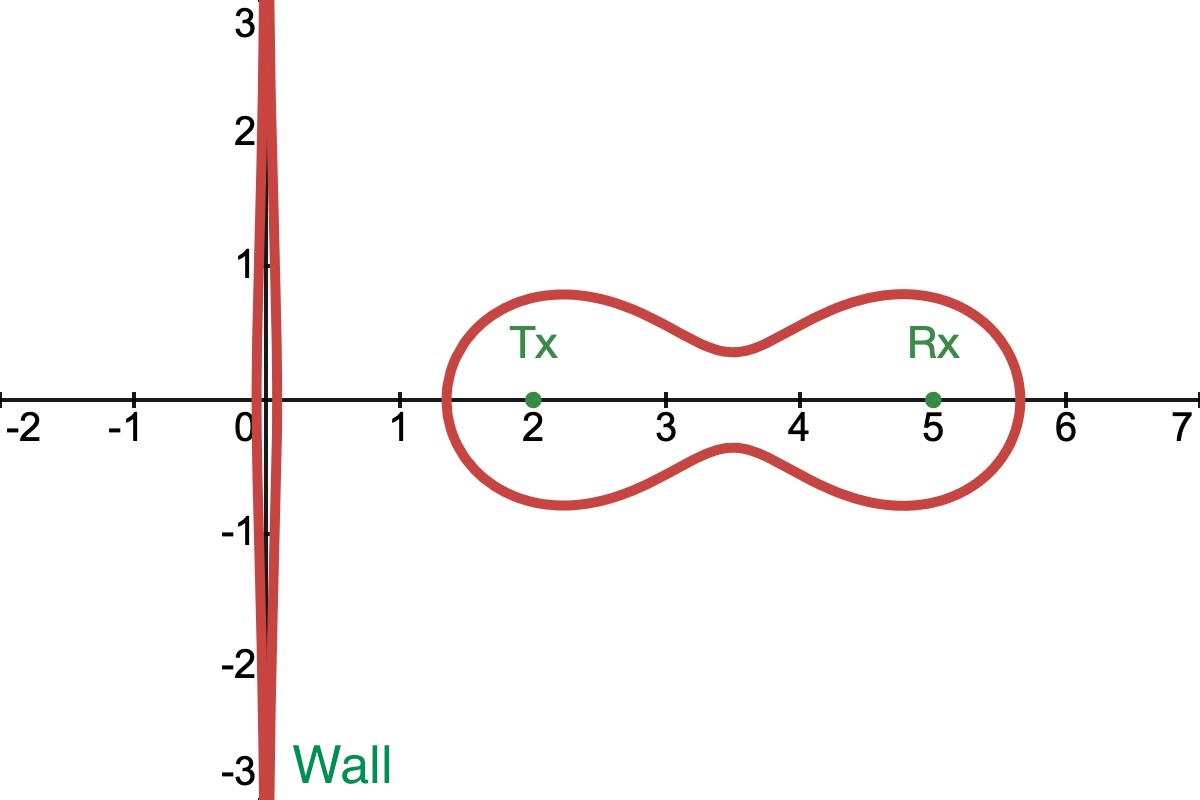}%
\label{wall_2}}
\subfloat[]{\includegraphics[width=2.1in]{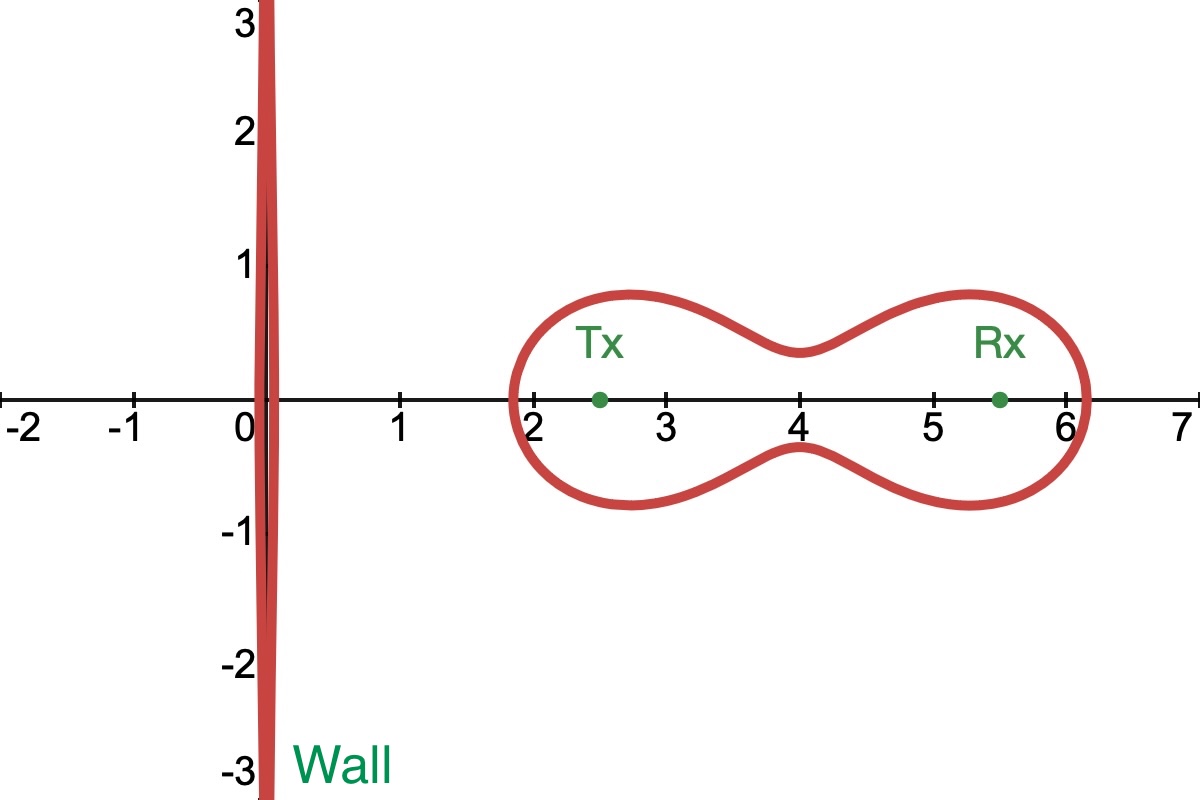}%
\label{wall_2.5}}
\caption{The sensing coverage boundary under different Wall-Tx distances. (a) Wall-Tx distance: 0.1 m. (b) Wall-Tx distance: 0.5 m. (c) Wall-Tx distance: 1 m. (d) Wall-Tx distance: 1.5 m. (e) Wall-Tx distance: 2 m. (f) Wall-Tx distance: 2.5 m.}
\label{bound_Wall_Tx}
\end{figure*}

We initially investigate the effect of the wall-device distance on the sensing boundary. Fig. \ref{bound_Wall_Tx} depicts the smoothened sensing boundary of a Wi-Fi sensing system with various distances between the wall and the transmitter, based on our analytical model in \eqref{SSNR_comb_simple}. 
It should be noted that we use an SSNR threshold of 2 dB to represent the sensing boundary, but the actual minimum SSNR required for sensing at the boundary may vary depending on the specific application, signal preprocessing techniques, and static power considerations. We maintain a fixed distance of 3 m between the Tx and Rx and increase the wall-transmitter distance from 0.1 m to 2.5 m. To ensure that only the wall-transmitter distance is altered, we simultaneously move both the Tx and Rx. As shown in Fig. \ref{wall_0.1}, Fig. \ref{wall_0.5}, and Fig. \ref{wall_1}, when the device is closer to the wall, the shape of the sensing coverage is affected by the wall and no longer like a Cassini oval \cite{placement_matters} with \eqref{SSNR_nowall_simple}. As the distance between the wall and the devices becomes larger, the impact of the wall on the sensing coverage diminishes. When the distance exceeds a certain threshold, such as 2 m shown in Fig. \ref{wall_2}, the shape and size of the sensing coverage closely resembles the sensing coverage model without a wall.

Based on this observation, we can conclude that the sensing coverage becomes larger when the devices are positioned closer to the wall within a certain range. However, when the device is too close to the wall, as depicted in Fig. \ref{wall_0.1}, the sensing coverage on the other side of the wall also increases. The expansion of the sensing coverage beyond the wall can lead to increased interference and adversely affect the performance, especially when there are other objects present on the other side of the wall. The underlying reason for this phenomenon is explained in the through-wall model proposed in \cite{wall_matters}. We will further evaluate and analyze this property in Section \ref{interference}. 

{\bfseries Factor II: Transmitter-Receiver Distance.}
\label{Section_Tx_Rx}

\begin{figure*}[!t]
\centering
\subfloat[]{\includegraphics[width=2.1in]{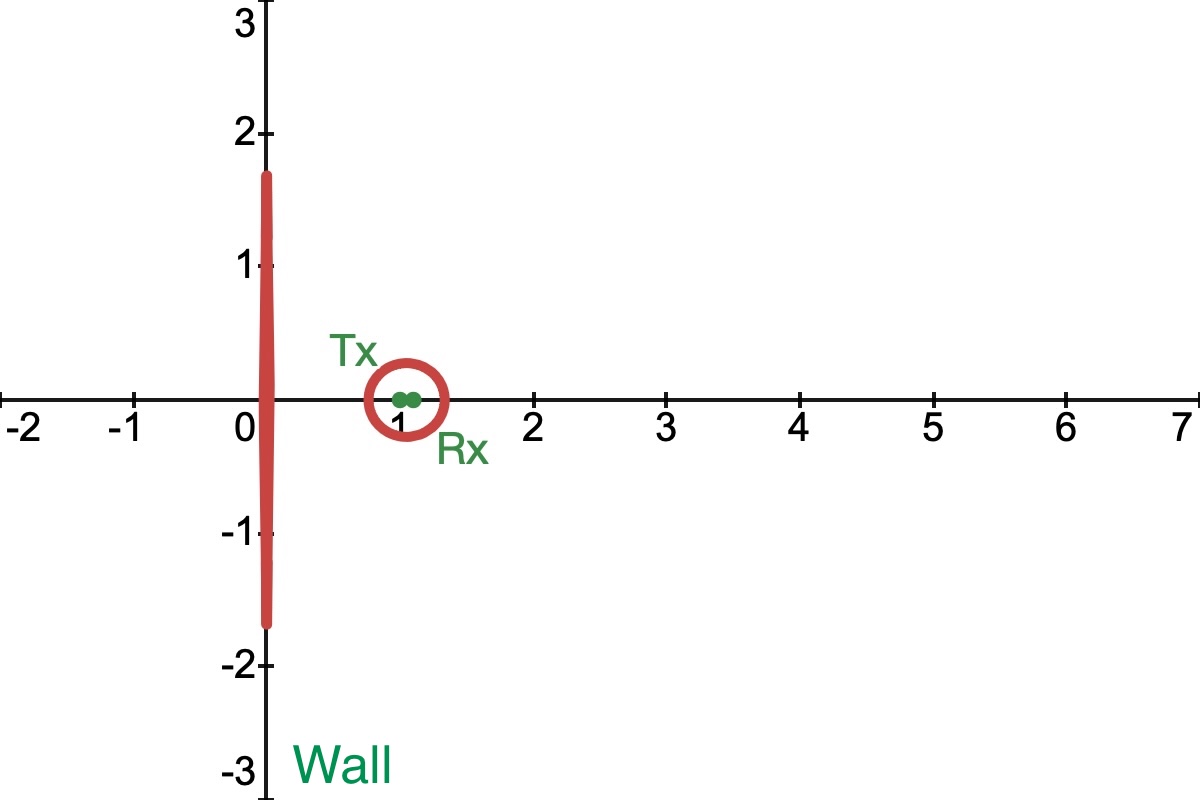}%
\label{TxRx_0.1}}
\subfloat[]{\includegraphics[width=2.1in]{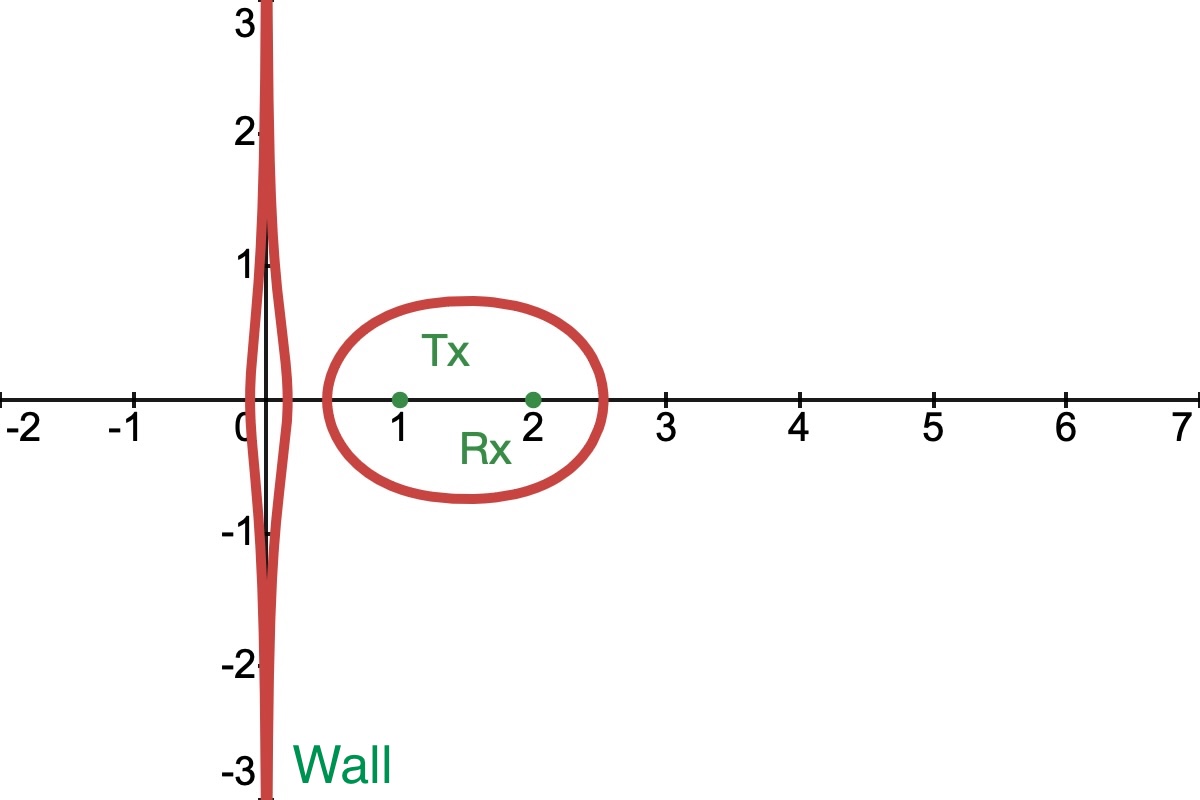}%
\label{TxRx_1}}
\subfloat[]{\includegraphics[width=2.1in]{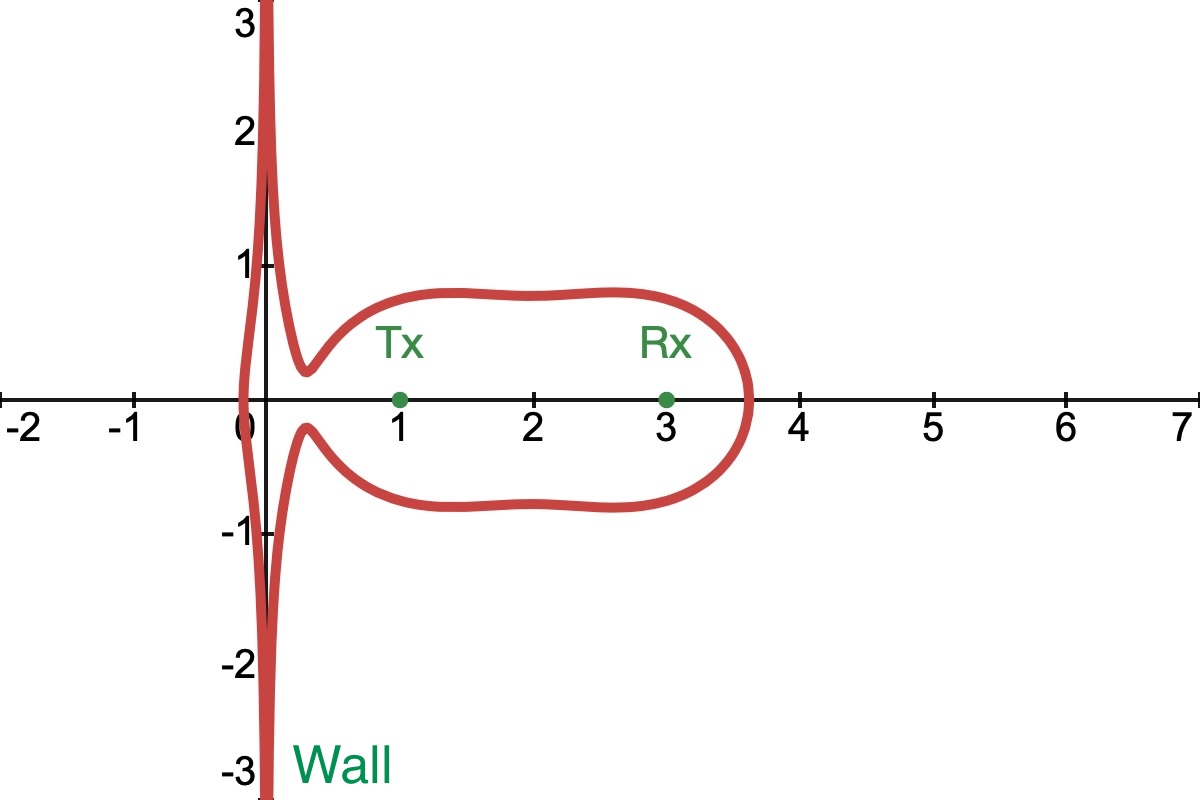}%
\label{TxRx_2}}

\subfloat[]{\includegraphics[width=2.1in]{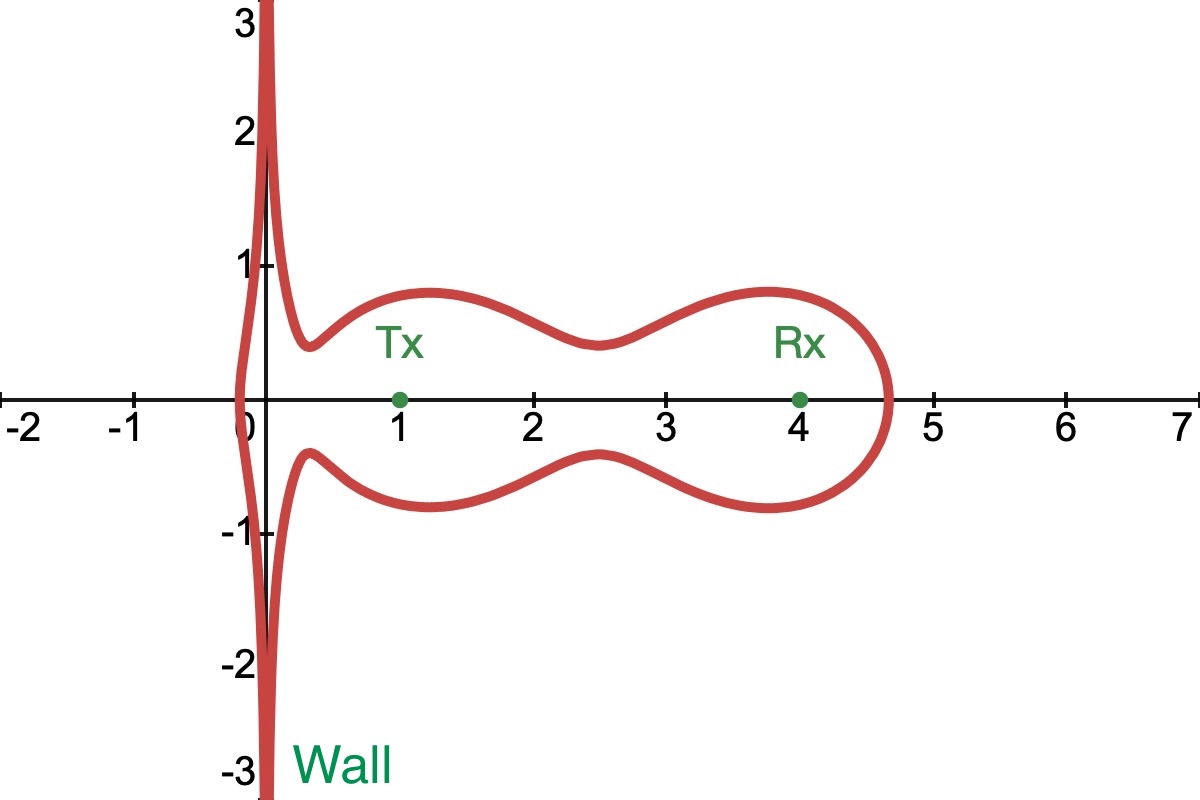}%
\label{TxRx_3}}
\subfloat[]{\includegraphics[width=2.1in]{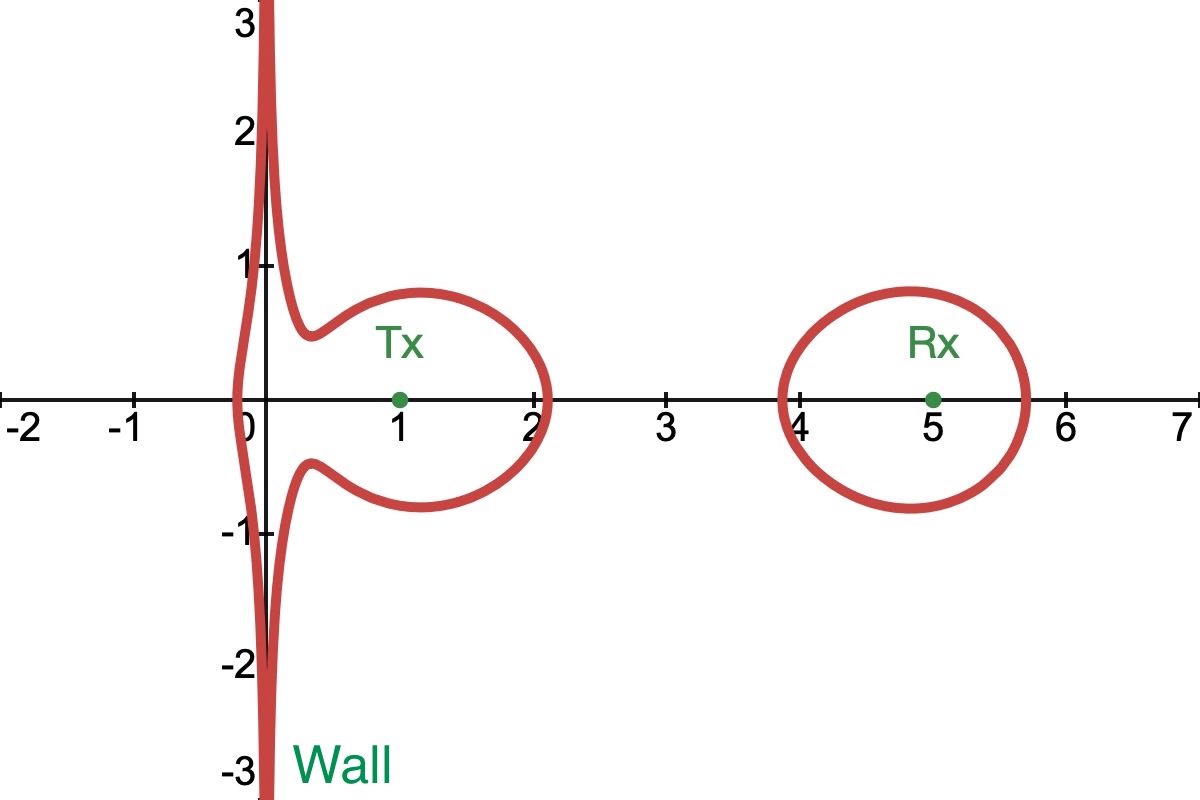}%
\label{TxRx_4}}
\subfloat[]{\includegraphics[width=2.1in]{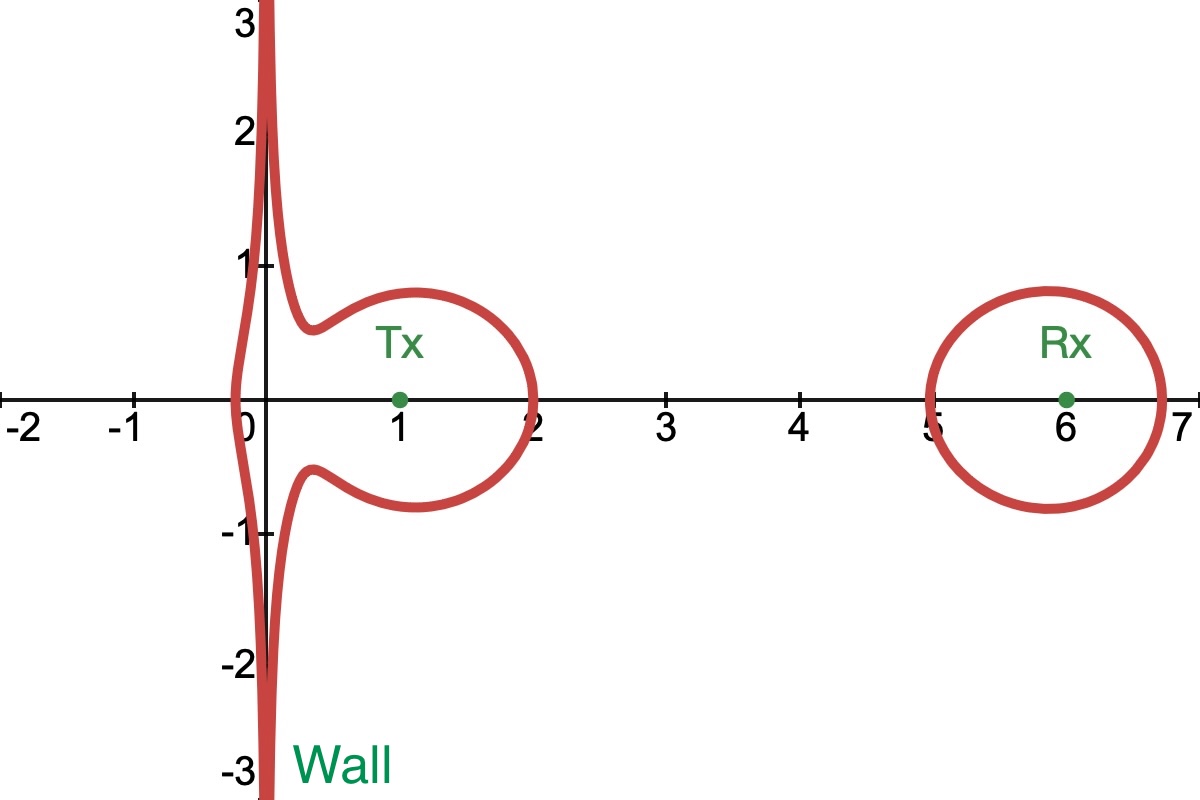}%
\label{TxRx_5}}
\caption{The sensing coverage boundary under different Tx-Rx distances. (a) Tx-Rx distance: 0.1 m. (b) Tx-Rx distance: 1 m. (c) Tx-Rx distance: 2 m. (d) Tx-Rx distance: 3 m. (e) Tx-Rx distance: 4 m. (f) Tx-Rx distance: 5 m.}
\label{bound_Tx_Rx}
\end{figure*}

To examine the impact of the Tx-Rx distance, we maintain the distance between the wall and transmitter at 1 m and move the receiver to increase the Tx-Rx distance from 0.1 m to 5 m. As illustrated in Fig. \ref{bound_Tx_Rx}, the corresponding sensing coverage initially increases and then decreases and separates into two distinct parts as the Tx-Rx distance increases. Specifically, when the distance is less than 3 m, the sensing coverage has an oval shape, and the coverage area expands as the Tx-Rx distance increases. However, when the distance exceeds 4 m, as depicted in Fig. \ref{TxRx_4}, the sensing coverage transforms into two circular regions surrounding the transceivers. Notably, the device situated closer to the wall exhibits a slightly larger sensing coverage compared to the device farther away from the wall. This discrepancy primarily arises from the increased dynamic power resulting from wall-reflected signals.

\subsection{Key Properties of the Indoor Sensing Coverage Model}

Based on the above theoretical analysis, we can summarize the key properties of Wi-Fi sensing coverage in indoor environments with respect to the wall-reflection effect as follows:

 \begin{enumerate}
     \item Whether walls have effects on the sensing coverage in Wi-Fi systems is dependent on the locations of devices relative to walls. When the distance between the device and the wall exceeds a threshold (e.g., 2 m in Fig. \ref{bound_Wall_Tx}), the sensing capability takes the shape of Cassini ovals. 
     \item When the distance between the wall and the device is short (e.g., 0.1 to 1 m in Fig. \ref{bound_Wall_Tx}), as the distance decreases, the sensing coverage expands both within the room and outside the room (on the other side of the wall).
     \item When the distance between the wall and the device is moderate (e.g., 1 - 2 m in Fig. \ref{bound_Wall_Tx}), the wall still has some effect on the sensing coverage inside the room, while the effect on the sensing coverage outside the room, beyond the walls, becomes negligible.
     \item When the transmitter-receiver distance is small (e.g., 0.1 - 3 m in Fig. \ref{bound_Tx_Rx}), the size and shape of the sensing coverage is a single small area. When the transmitter-receiver distance is large (e.g., $>$ 4 m in Fig. \ref{bound_Tx_Rx}), the size and shape of the sensing coverage becomes several separated small areas surrounding the wall, transmitter, and receiver.
 \end{enumerate}

\section{Evaluation}
\label{Section_VI}

In this section, our focus is on validating the key properties of the indoor sensing coverage model through two specific applications. First, respiratory monitoring is employed to evaluate the effectiveness of the SSNR-based model regarding its sensing capabilities. Second, stationary crowd counting is utilized to assess the sensing coverage properties of the proposed model. The experiments conducted on these applications provide a foundation for our subsequent analysis, allowing us to validate the reliability and accuracy of our findings. Leveraging the properties obtained from these experiments, we then utilize them to guide the system design of stationary crowd counting with various scenarios.

\begin{figure}[h]
  \centering
  \includegraphics[width=0.8\linewidth]{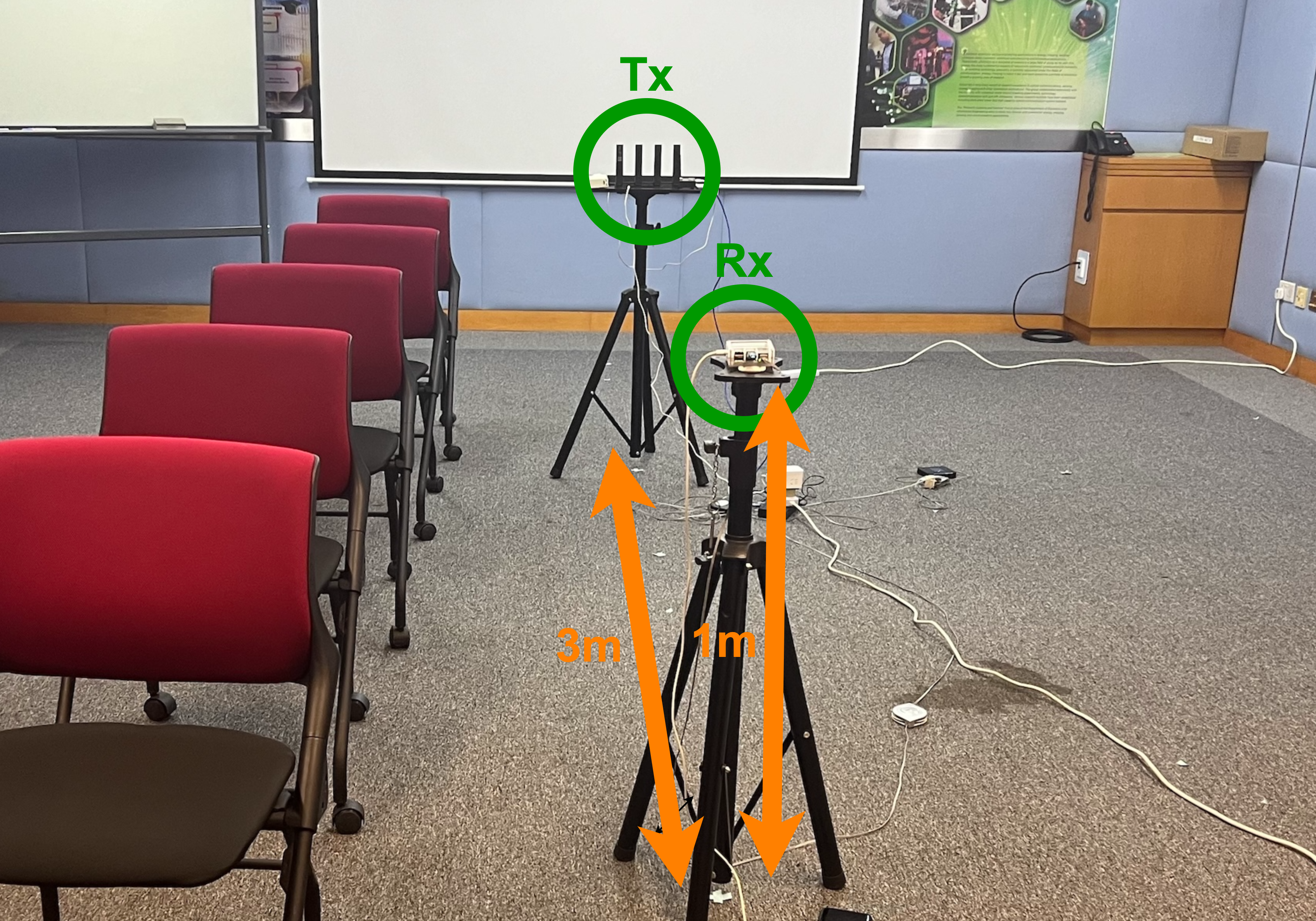}
  \caption{Hardware setup in a real scenario.}
  \label{CD634_real}
\end{figure}

{\bfseries Hardware Setup.} 
In the conducted experiments, a router with omni-directional antennas was used as the transmitter, while a Raspberry Pi equipped with the Nexmon CSI Extraction Tool \cite{free_your_csi} served as the receiver. The employed tool is able to support 802.11a/g/n/ac transmissions in both the 2.4 and 5 GHz bands, with a maximum bandwidth of 80 MHz. To maximize the available information, we utilized the 80 MHz bandwidth in the 5 GHz band, capturing 242 subcarriers after removing nulls. Since the Raspberry Pi has a single antenna and its Wi-Fi chip operates on a single core, each Pi collected a single CSI matrix. The sampling frequency was set to 1000 Hz. In order to ensure stability, the transceivers were mounted on tripods at a height of 1 m. Fig. \ref{CD634_real} provides an example of the experimental setup in a meeting room.

{\bfseries Evaluation Setup.}
The key of this paper is to demonstrate the improvement contributed by the effect of walls, and the relationship between the proposed model and the performance of sensing applications in real scenarios. We conduct experiments in several representative indoor environments with different room layouts and wall materials to provide initial validation of the robustness and general applicability of our findings. In the context of respiratory monitoring, our methodology integrates signal processing techniques such as the Hampel filter \cite{hampel_filter_2016}, Savitzky-Golay filter \cite{sg_filter}, and peak detection algorithms to compute respiratory rates in breaths per minute (bpm). A reference respiration rate is acquired from a Neulog Respiration Monitor Belt logger sensor NUL236. The Mean Absolute Error (MAE) metric \cite{mae2005} is employed to evaluate the application performance, defined as
\begin{equation}
\mathrm{MAE} = \frac{1}{M}\sum_{m=1}^{M}\left|\hat{b}_m - b_{\mathrm{ref}}\right|~\text{(bpm)},
\end{equation}
where $\hat{b}_m$ and $b_{\mathrm{ref}}$ denote the estimated and reference respiration rates, respectively.

For stationary crowd counting accuracy evaluations, we leverage traditional machine learning models, namely Support Vector Machines (SVM), Random Forest (RF), and K-Nearest Neighbors (KNN). The accuracy metric is defined as
\begin{equation}
\mathrm{Acc} = \frac{1}{M}\sum_{m=1}^{M}\mathbb{I}(\hat{c}_m = c_m),
\end{equation}
where $\mathbb{I}(\cdot)$ equals 1 when the predicted label $\hat{c}_m$ matches the ground truth $c_m$, and 0 otherwise. The preprocessing procedures and model parameters employed in these models remained consistent with our previous research \cite{WCNC_my}. Although more advanced signal processing and deep learning methods have been reported to achieve accuracy rates exceeding 99\% \cite{crowd_dl99_2025}, such saturation would obscure the relative differences across transceiver placements. Therefore, we intentionally adopt simple and well-validated models to ensure controlled and meaningful comparisons, allowing a clear evaluation of the sensing coverage enhancements introduced by wall proximity.

\subsection{Evaluating Sensing Capability of the Proposed Model through Respiratory Monitoring}
\label{subsection_RM}

To validate the properties of the proposed model concerning sensing capability, we conducted experiments specifically tailored for respiration monitoring applications. We compared the simulated SSNR, the real-scenario SSNR, and the MAE results in various scenarios.

For the real-scenario SSNR, the dynamic power is computed based on the disparity between the CSI amplitude and the averaged amplitude within the observation window. The interference power is determined as the difference before and after filtering the CSI amplitude \cite{placement_matters}. As illustrated in Fig. 9, when the device is positioned close to the wall (0.1 m, Fig. 9a), the CSI amplitude exhibits a high-quality respiratory pattern with minimal noise and a peak-to-peak amplitude exceeding 100. In contrast, at a Wall–Tx distance of 1 m (Fig. 9c), the signal becomes significantly noisier and the amplitude variation drops below 20. This degradation in signal clarity and amplitude range results in a lower SSNR value, validating the advantage of wall-proximity deployment for respiration-based sensing.

\begin{figure*}[!t]
\centering
\subfloat[]{\includegraphics[width=2.1in]{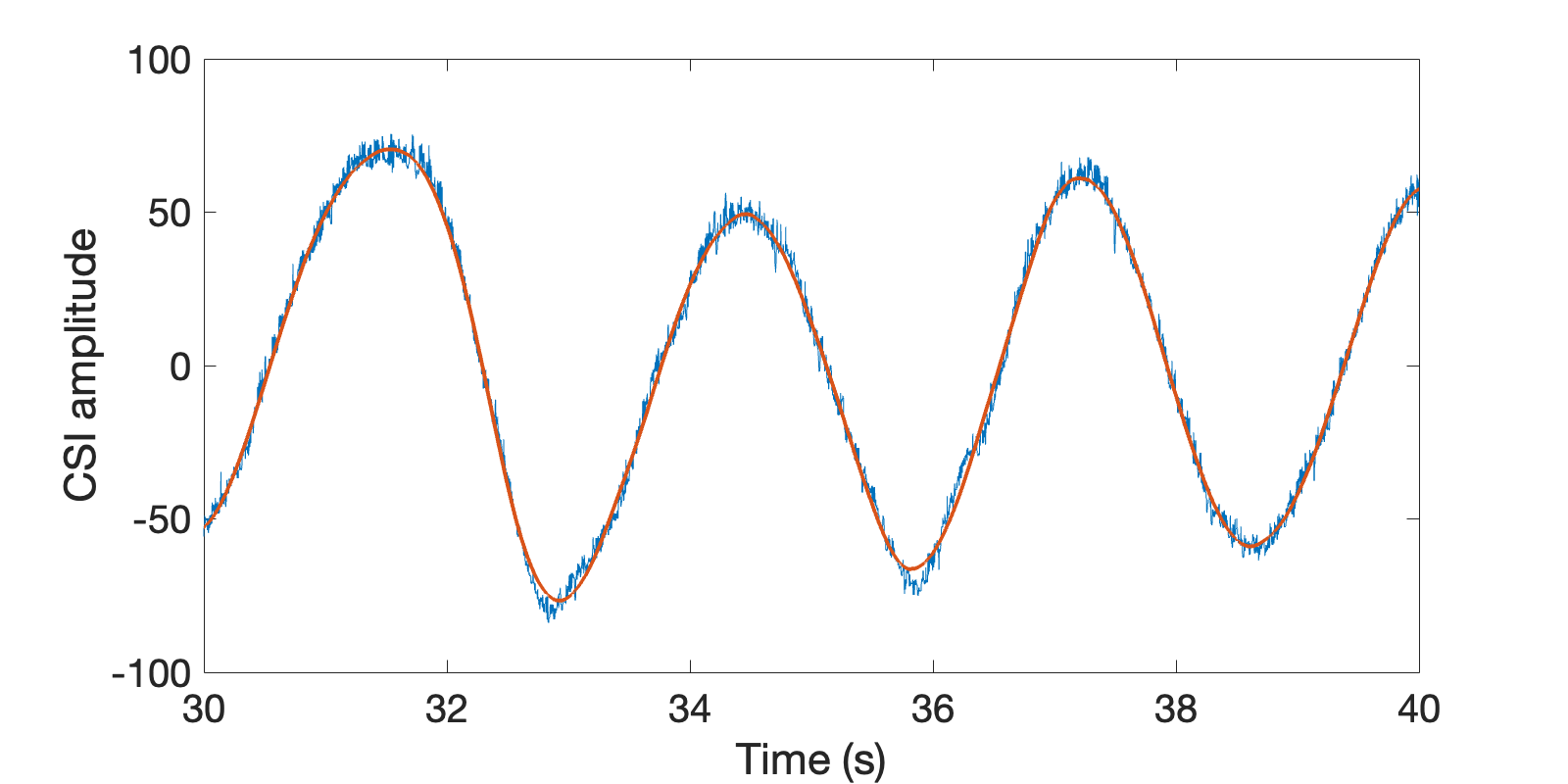}%
\label{RR_0.1}}
\subfloat[]{\includegraphics[width=2.1in]{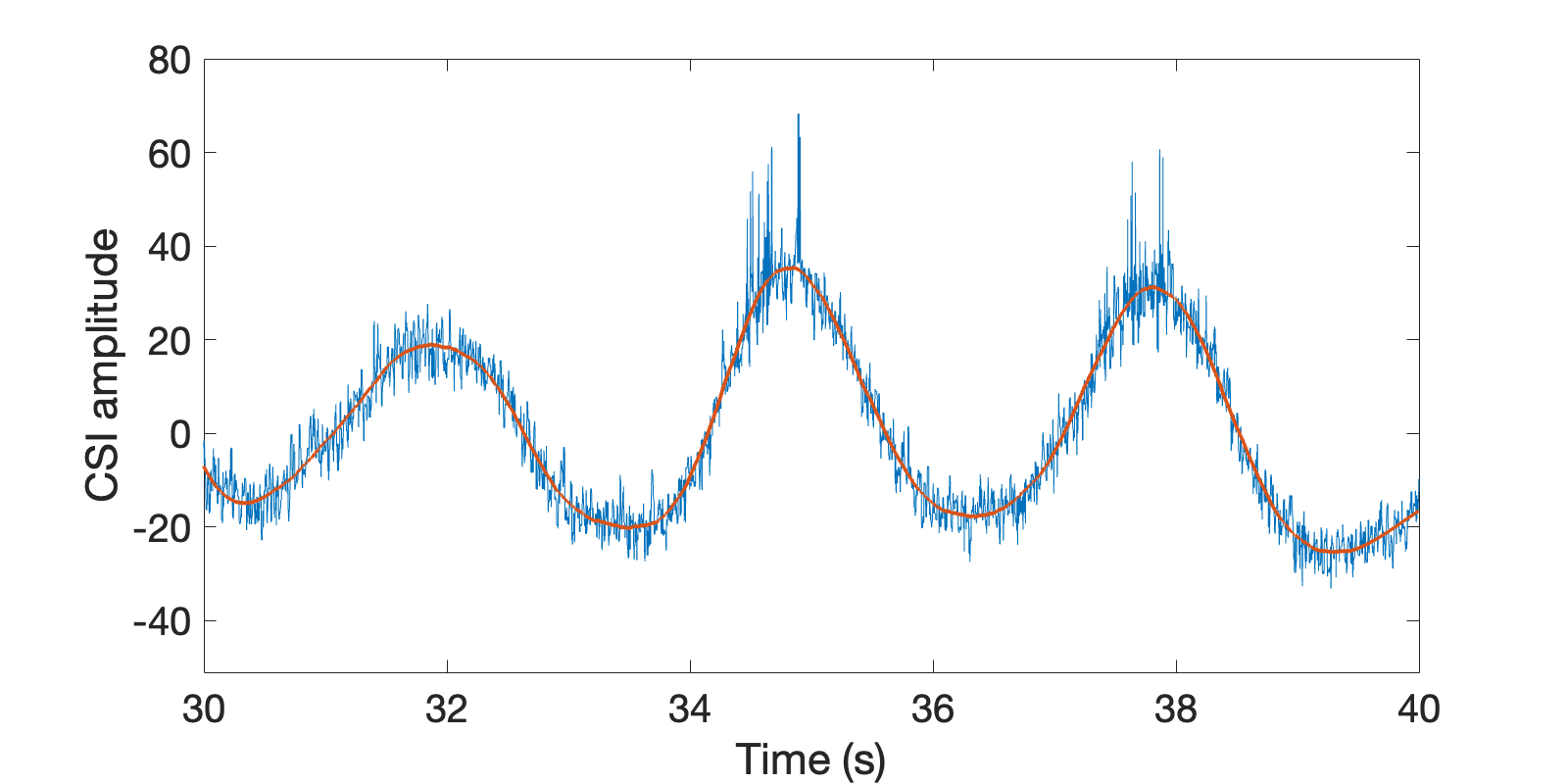}%
\label{RR_0.5}}
\subfloat[]{\includegraphics[width=2.1in]{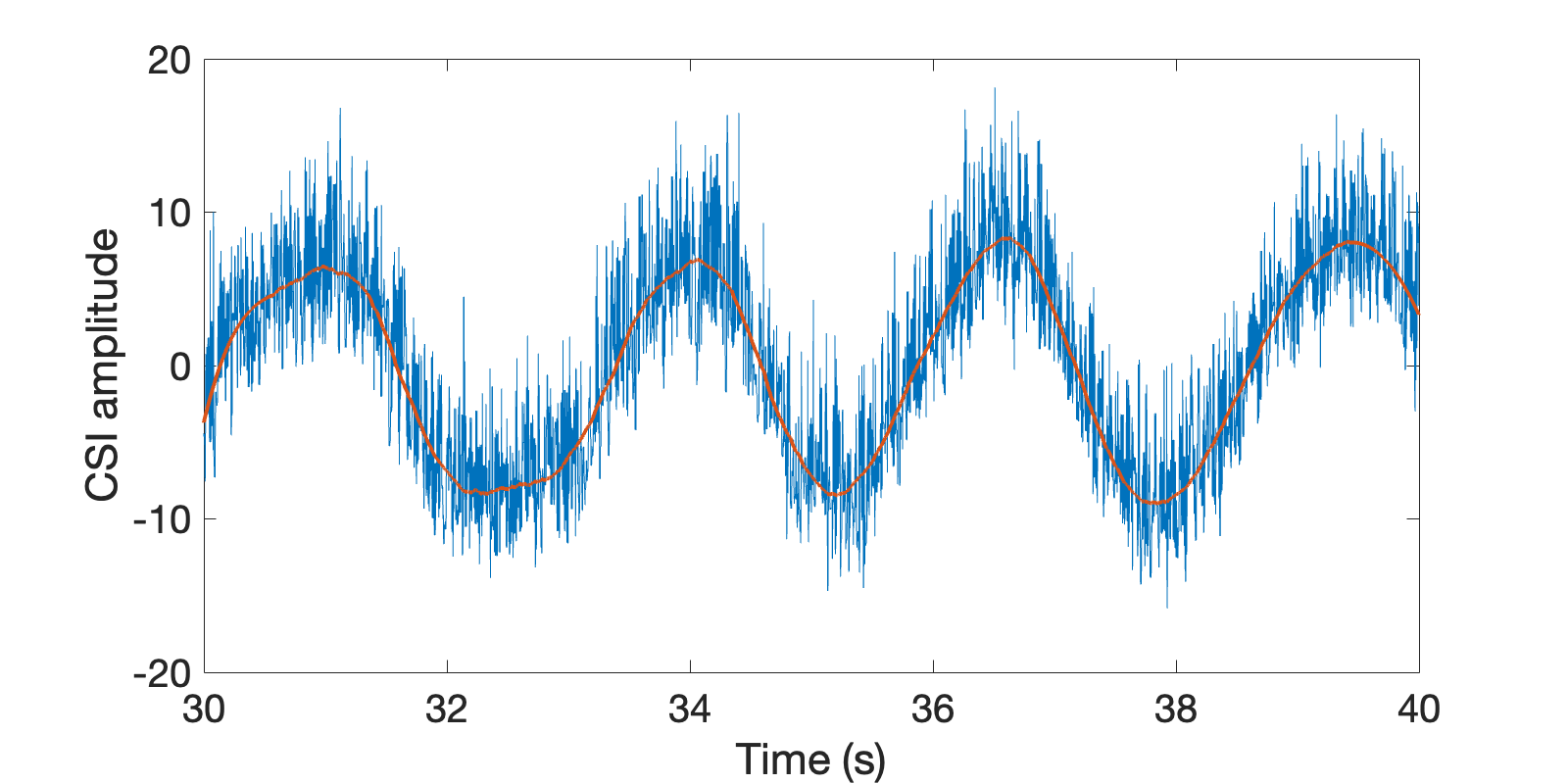}%
\label{RR_1}}

\caption{The respiratory signal under different Wall-Tx distances. (a) Wall-Tx distance: 0.1 m. (b) Wall-Tx distance: 0.5 m. (c) Wall-Tx distance: 1 m.}
\label{RR_example}
\end{figure*}

{\bfseries Experiment 1: Verify the effectiveness of using simulated SSNR to assess the SSNR measured in real scenarios}
\label{RR_Ex1}

\begin{figure}[h]
  \centering
  \includegraphics[width=0.7\linewidth]{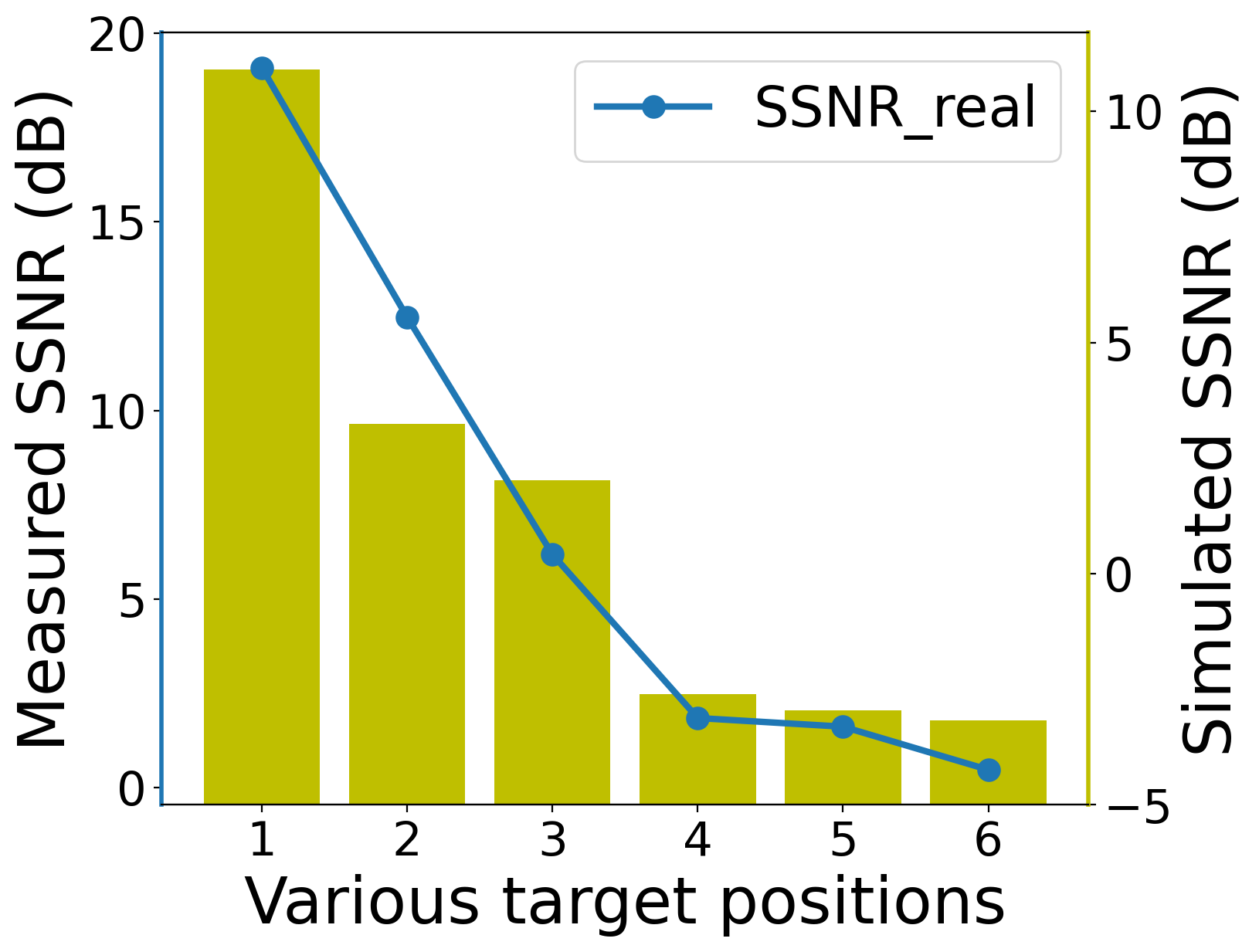}
  \caption{Comparison between simulated and measured SSNR at different target positions. Bars: simulated SSNR; line: measured SSNR.}
  \label{RR_SSNR_sim_real}
\end{figure}

In this experiment, the data with a single target respirating at various locations is examined. Fig. \ref{RR_SSNR_sim_real} illustrates the comparison between the simulated SSNR and the SSNR derived from real-world data. It is evident that the SSNR in the real scenario aligns with the simulated SSNR by a scale factor. This factor is contingent upon parameters such as device power, transceiver deployment, and environmental conditions. In our simulated model, we normalized the effect of these parameters, as they do not significantly impact the correlation between wall, device, target distances, and sensing coverage. These results validate the effectiveness of our proposed SSNR-based model in assessing sensing capability and coverage.  

{\bfseries Experiment 2: Verify the effect of the wall-device distance without interference}
\label{RR_Ex2}

\begin{figure}[h]
  \centering
  \includegraphics[width=0.9\linewidth]{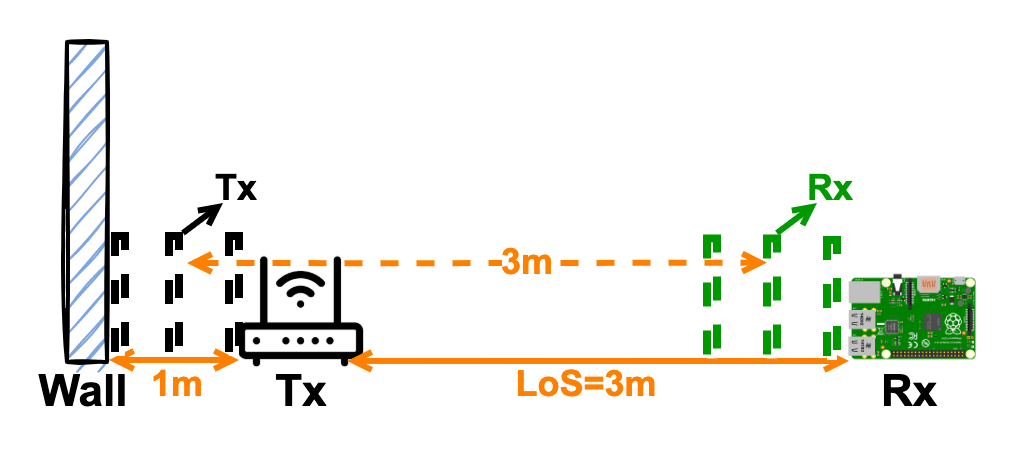}
  \caption{Verify the effect of the wall-device distance by moving transceivers at the same time.}
  \label{Wall_Tx_0_1m}
\end{figure}

\begin{figure}[ht]
  \centering
  \subfloat[]{%
    \includegraphics[width=0.48\linewidth]{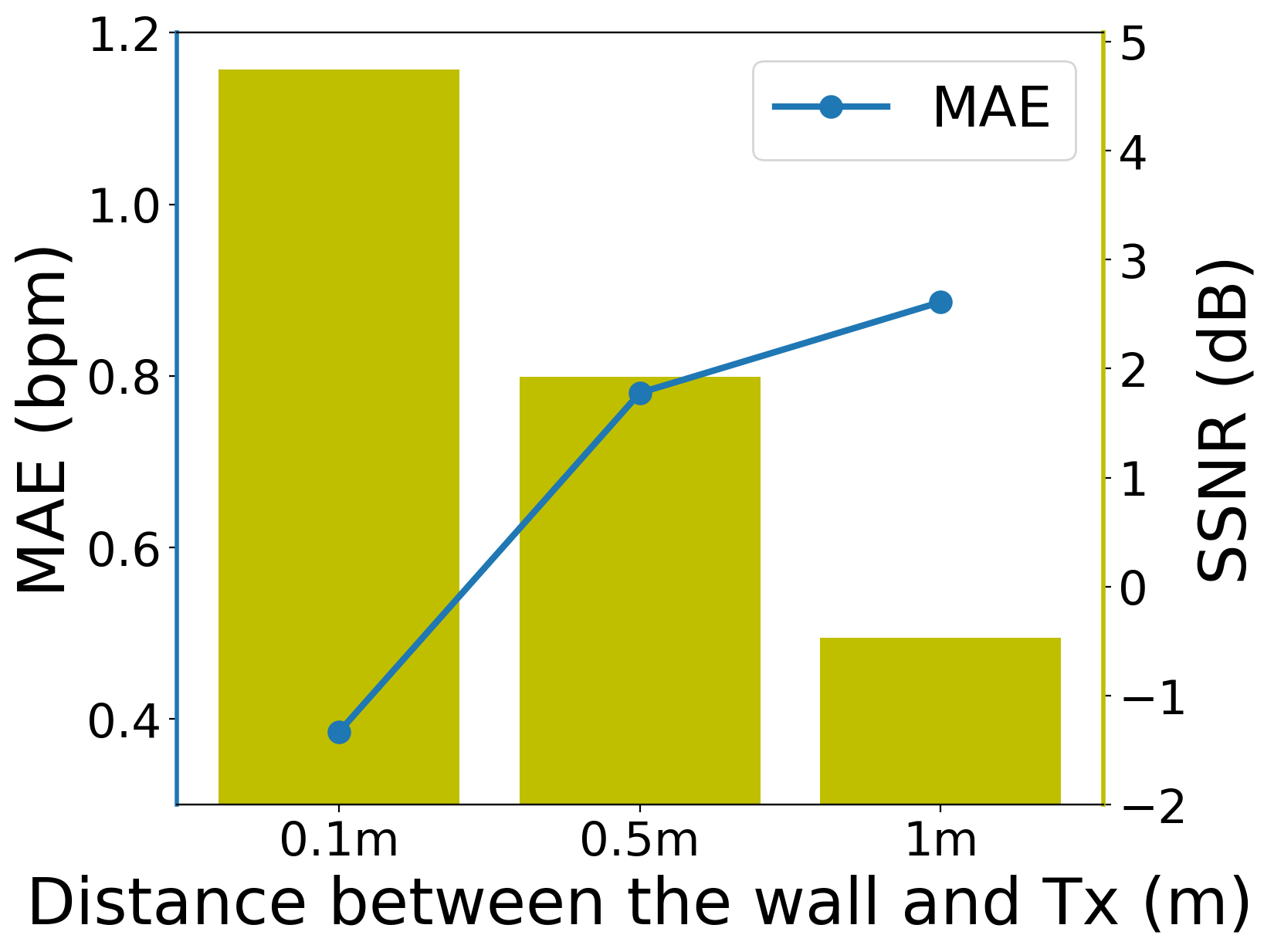}
    \label{RR_SSNR_real_MAE}
  }
  \hfill
  \subfloat[]{%
    \includegraphics[width=0.48\linewidth]{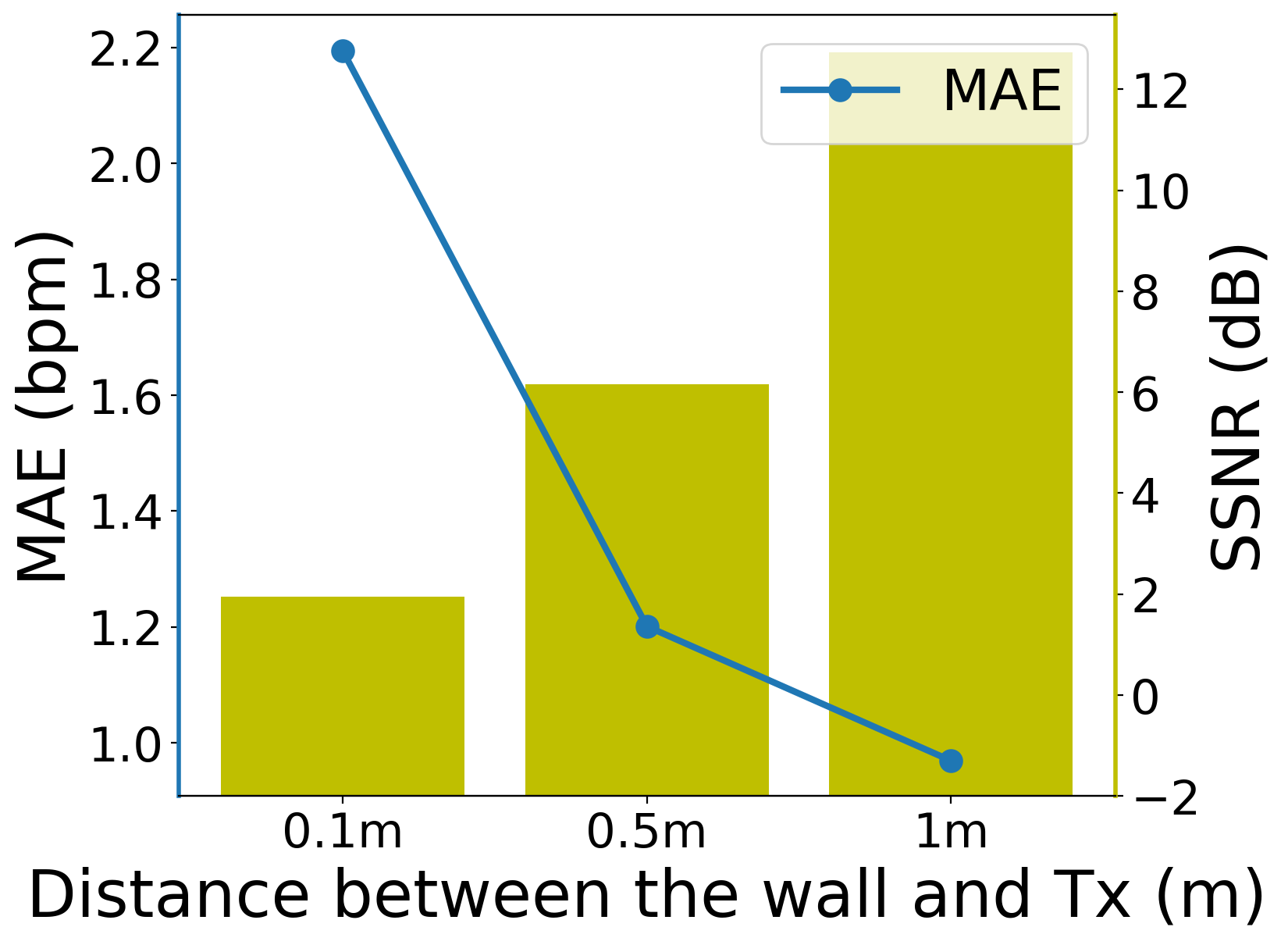}
    \label{RR_SSNR_real_MAE_noise} 
  }
  \caption{Evaluation results of the effect of the wall–device distance on sensing performance. (a) Without interference. (b) With interference. Bars: SSNR; lines: MAE.}
  \label{fig:RR_SSNR_compare} 
\end{figure}

In this experiment, we aim to verify the impact of the wall-device distance on the sensing capability. We position the transmitter at distances of 0.1 m, 0.5 m, and 1 m away from the wall, as depicted in Fig. \ref{Wall_Tx_0_1m}. The distance between the Tx and Rx is fixed at 3 m, with the Rx moving simultaneously with the Tx. Fig. \ref{RR_SSNR_real_MAE} presents the MAE (line chart) obtained for the three cases, along with the corresponding SSNR value (bar chart). We observe that the experimental result aligns well with the simulation trend. Specifically, as the distance between the wall and the device decreases within a limited range, the sensing capability increases, and the MAE decreases. This finding confirms the relationship between the wall-device distance and the extent of the sensing capability, as predicted by the simulation property described as Factor I in Section \ref{Section_Wall_Tx}.

{\bfseries Experiment 3: Verify the effect of the wall-device distance with interference}
\label{RR_Ex3}

The aim of this experiment is to verify the impact of the wall-device distance on sensing area expansion and its potential negative effect when interferers are present within this expanded area. The experimental setup mirrors that of Experiment 2, with the addition of an interferer seated approximately 0.5 meters outside the wall and capable of upper-body movement. In Fig. \ref{RR_SSNR_real_MAE_noise}, the green bars depict SSNR values in real scenarios, while the blue line represents the MAE for the target within the room.

As the transmitter moves closer to the wall, the SSNR decreases, indicating significant interference in this case. The trend observed in Fig. \ref{RR_SSNR_real_MAE_noise} demonstrates that as the wall-device distance increases, the MAE decreases. This phenomenon occurs because a shorter wall–device distance not only strengthens the main-wall reflection within the room but also increases the transmission of signals through the wall, allowing external motions to couple into the sensing channel. While the present study primarily focuses on wall-reflection effects within the room, this observation highlights that excessive proximity to the wall may also amplify cross-wall interference through transmitted components.

\subsection{Evaluating Sensing Coverage of the Proposed Model through Stationary Crowd Counting}
\label{cc_1}

To validate the sensing coverage aspects of the proposed model, we conducted experiments tailored specifically for stationary crowd counting applications. The experimental setup in Sections \ref{cc_1} and \ref{cc_2} included five participants of varying heights and weights, consisting of one female and four males, aged 24 to 26 years. Each experiment consisted of three trials, with participants positioned at a total of 15 distinct locations distributed randomly throughout the entire space. In these experiments, we varied the wall-device distance while keeping other factors constant, such as the participants' positions and the total number of participants. Since crowd counting involves multiple static targets simultaneously, the SSNR metric is not applicable in this scenario. Instead, we validate the model and its simulated sensing coverage by correlating the estimated sensing areas with the counting accuracy. It should be noted that the simulated sensing area derived from the SSNR-based model serves as a theoretical guideline rather than a strict predictor of recognition accuracy. Higher accuracy indicates better performance and implies a larger effective sensing coverage under ideal conditions. However, this relationship may not always hold in practice, as larger coverage could also introduce greater vulnerability to external interference. This trade-off will be further examined in subsequent experiments. While the real sensing area may extend beyond the simulated boundary due to advanced signal preprocessing, the overall trends observed across varying device positions should still align with the model predictions.

{\bfseries Experiment 1: Verify the effect of the wall-device distance without interference}

\begin{figure}[ht]
  \centering
  \subfloat[]{%
    \includegraphics[width=0.48\linewidth]{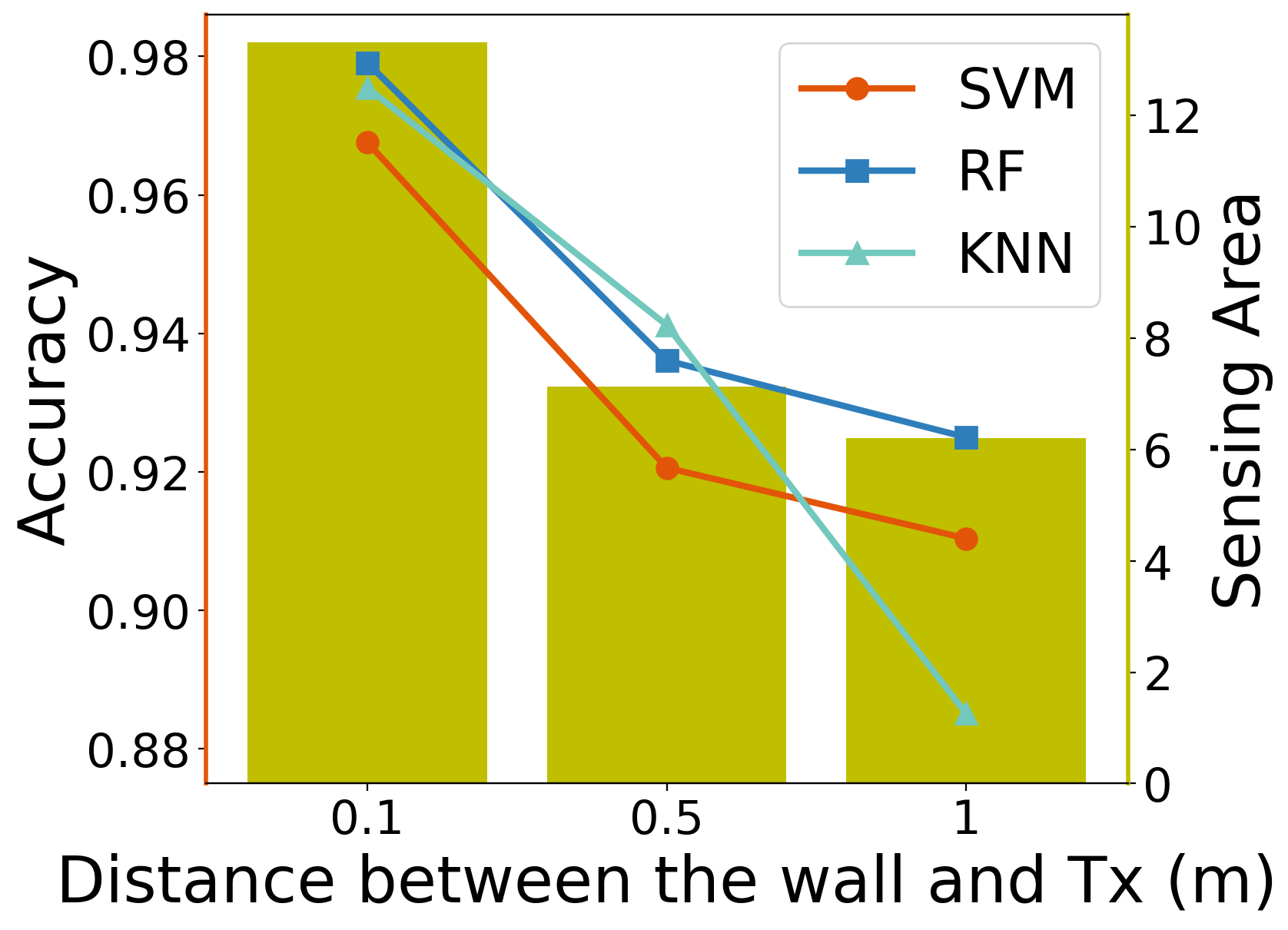}
    \label{Wall_Tx_0_1m_no_noise} 
  }
  \hfill
  \subfloat[]{%
    \includegraphics[width=0.48\linewidth]{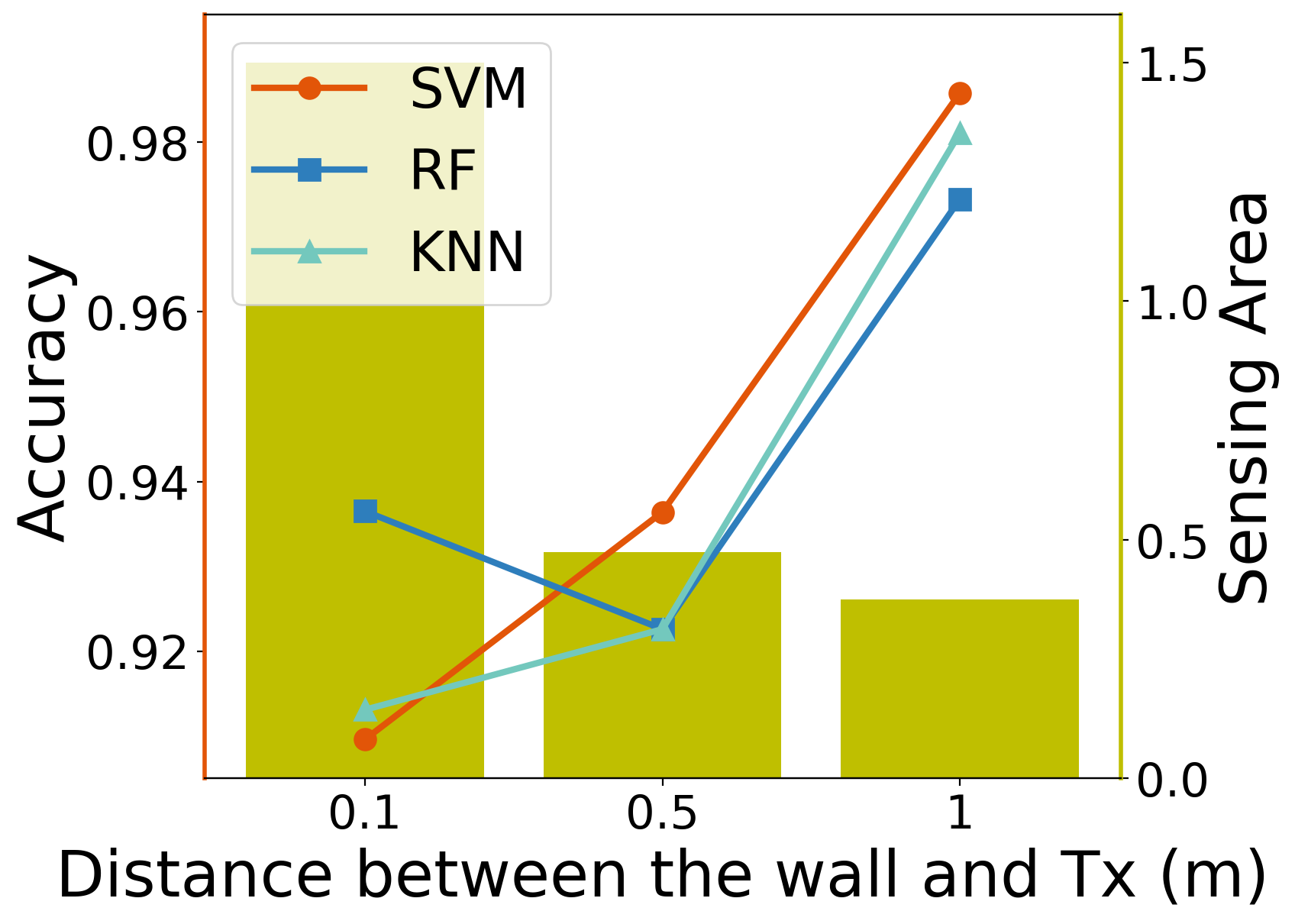}
    \label{Wall_Tx_0_1m_noise}  
  }
  \caption{Evaluation results of the effect of the wall-device distance on sensing performance. (a) Without interference. (b) With interference. Bars: sensing area; lines: accuracy.}
  \label{fig:Wall_Tx_compare}
\end{figure}

In this experiment, we aim to verify the impact of the wall-device distance on the sensing coverage, as discussed in Section \ref{Section_Wall_Tx}. The device deployment settings are the same as in Section \ref{RR_Ex2}, as depicted in Fig. \ref{Wall_Tx_0_1m}. Fig. \ref{Wall_Tx_0_1m_no_noise} presents the accuracy results (line charts) obtained for the three cases, along with the corresponding simulation-based sensing areas (bar chart). We observe that the experimental results across all three models align well with the simulation trend. Specifically, as the distance between the wall and the device decreases within a limited range, the indoor sensing area increases. This finding confirms the relationship between the wall-device distance and the extent of the sensing coverage, consistent with the simulated property referred to as Factor I in Section \ref{Section_Wall_Tx}.

{\bfseries Experiment 2: Verify the effect of the wall-device distance with interference}
\label{interference}

As mentioned in Section \ref{section_reflectwall}, when the device is close to the wall, the sensing area on the other side of the wall also increases, which will have a negative effect on indoor sensing applications. This experiment is designed to verify this property. The experiment setup is the same as Experiment 1 except that additional people are positioned outside the room, adjacent to the wall, introducing external interference. In Fig. \ref{Wall_Tx_0_1m_noise}, the green bars represent the sensing areas beyond the wall and the three lines show the accuracy results across all three models for people counting within the room. As the Tx moves closer to the wall, the sensing area on the other side of the wall extends, which indicates a large interference area in this experiment. As can be seen from Fig. \ref{Wall_Tx_0_1m_noise}, with a decrease in the wall-device distance, the accuracy also decreases. This trend verifies the negative effect of a short wall-device distance when there are interferers beyond the wall.

{\bfseries Experiment 3: Verify the effect of the transmitter-receiver distance}

\begin{figure}[h]
  \centering
  \includegraphics[width=0.9\linewidth]{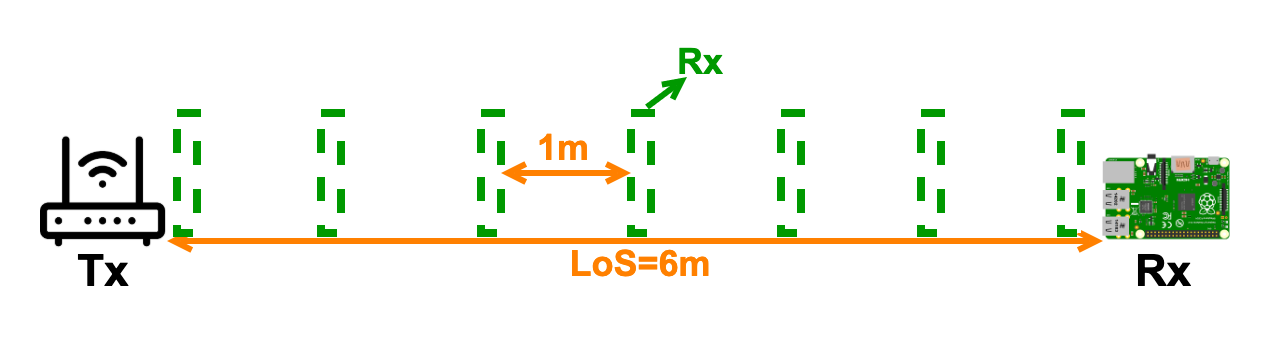}
  \caption{Verify the effect of the Tx-Rx distance by placing the transceivers at different locations.}
  \label{Tx_Rx_0_6m}
\end{figure}

\begin{figure}[h]
  \centering
  \includegraphics[width=0.9\linewidth]{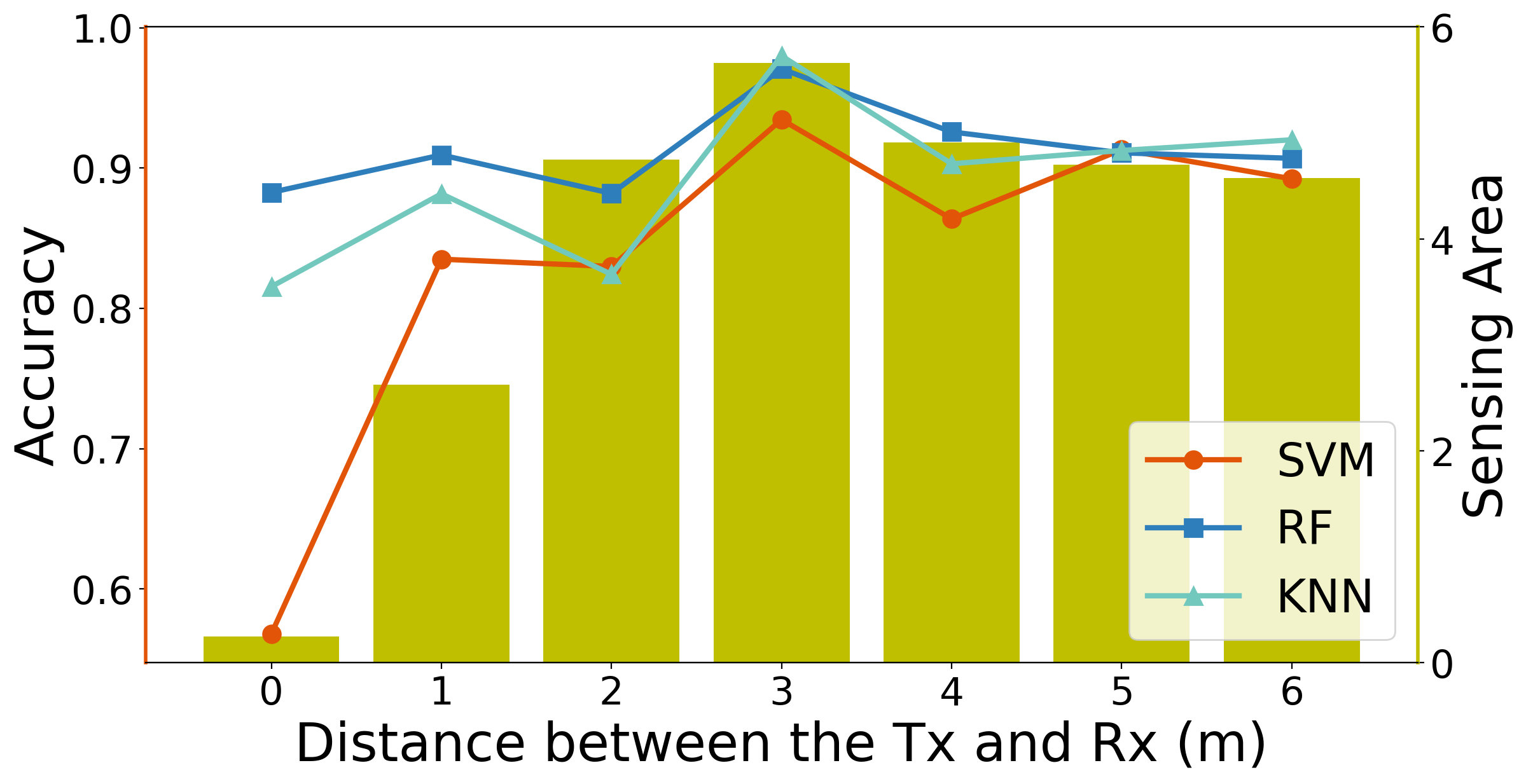}
  \caption{Evaluation results of the effect of the Tx-Rx distance. Bars: sensing area; lines: accuracy.}
  \label{Tx_Rx_various_distances}
\end{figure}

To evaluate the effect of the transmitter-receiver distance on the proposed model, we conducted experiments in a meeting room with a fixed distance of 1.2 m between the wall and the Tx. As depicted in Fig. \ref{Tx_Rx_0_6m}, we varied the distance between the transceivers from 0 m to 6 m at a step size of 1 m. The results, shown in Fig. \ref{Tx_Rx_various_distances}, indicate that the accuracy initially increases and then slightly decreases with the increased distance between the Tx and Rx. This trend is consistent across all the three models, and it aligns with the simulated sensing area presented in the bar chart. Notably, an optimal distance of 3 m emerged, resulting in the largest sensing area and the highest accuracy for stationary crowd counting in this particular environment. These experimental results serve to validate the property of the proposed model referred to as Factor II in Section \ref{Section_Tx_Rx}.

While the phenomena observed in Experiment 3 could also be explained by the existing model \cite{placement_matters}, it fails to capture the key observations from Experiments 1 and 2, as the previous model does not consider the influence of walls on dynamic sensing capabilities. In contrast, our model uniquely demonstrates the effects of wall-device distance, offering a more comprehensive understanding of deployment strategies in real-world scenarios. This deeper understanding suggests that while closer proximity to the wall can improve sensing coverage, it may be counterproductive in environments with potential interference sources. Therefore, deployment decisions should balance sensing range expansion with environmental robustness, particularly in complex indoor settings.

\subsection{Boosting the Sensing Coverage of Stationary Crowd Counting}
\label{cc_2}

In this section, we conduct experiments with stationary crowd counting on various scenarios to evaluate the effectiveness of utilizing the properties outlined in this paper to expand the sensing coverage. We categorize the deployment strategies into two types: i) Ensuring that one of the transceivers remains in close proximity to the wall while maintaining an optimal distance between the transceivers; and ii) Positioning both devices in close proximity to the walls.

{\bfseries Case 1: Scenarios where only one device is positioned close to the wall}

\begin{figure}[h]
\centering
\subfloat[]{\includegraphics[width=1.7in]{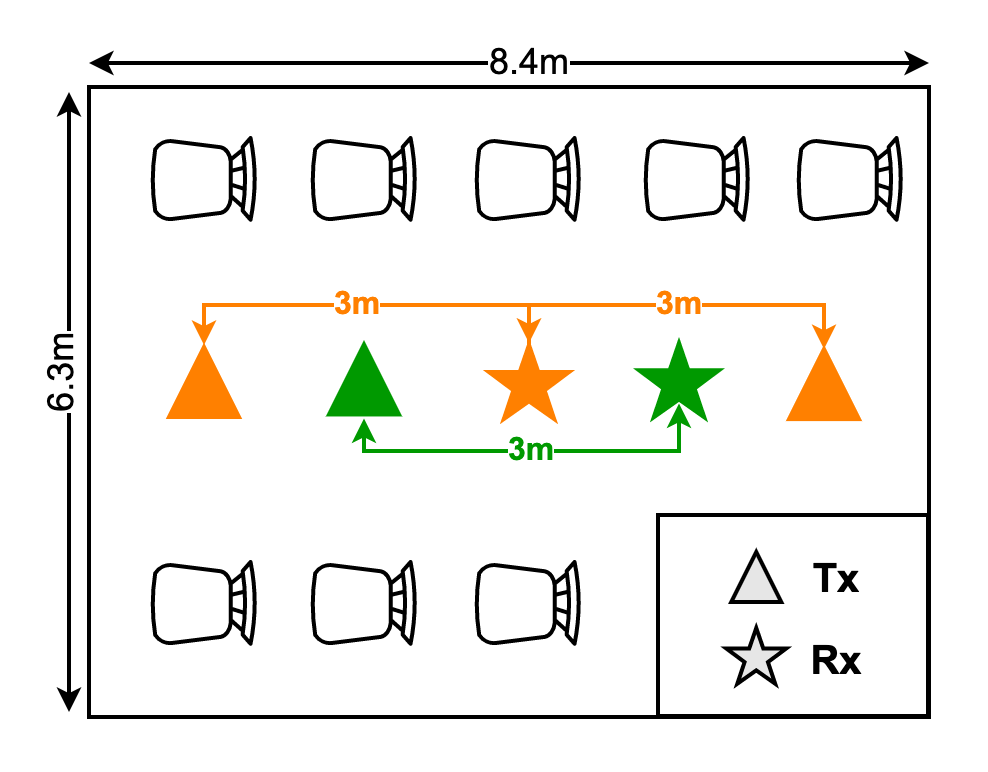}%
\label{CD634}}
\subfloat[]{\includegraphics[width=1.7in]{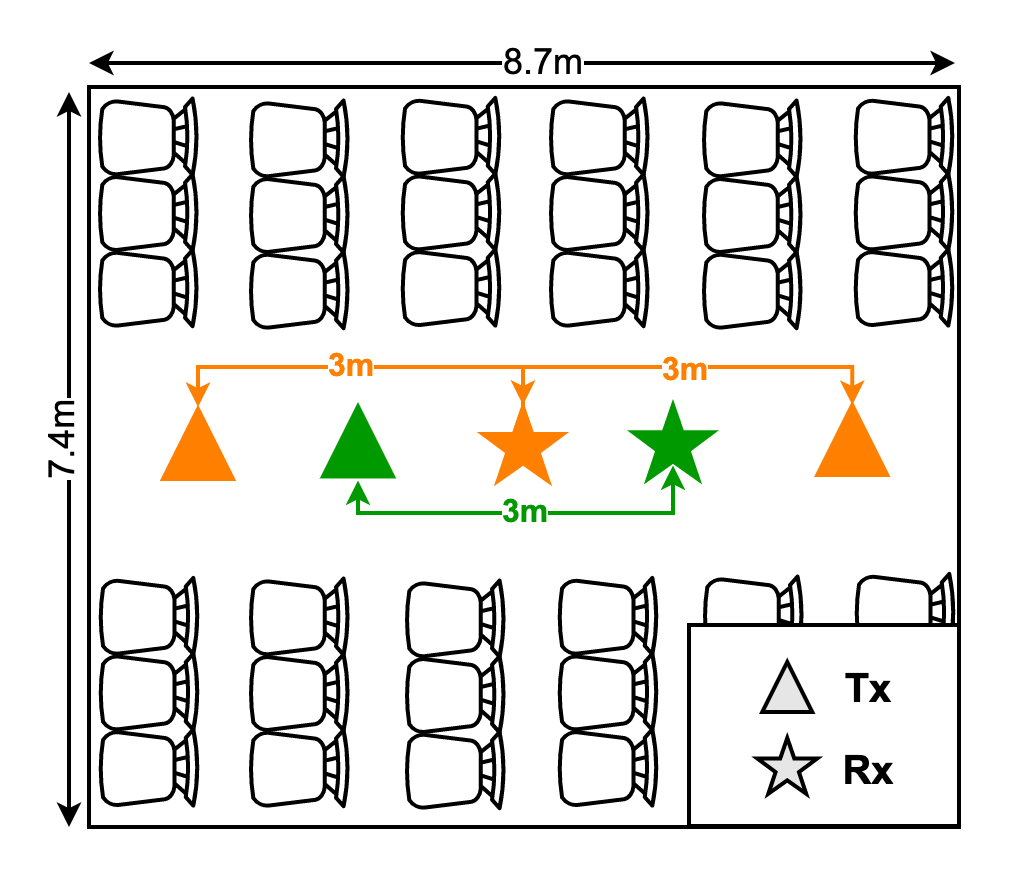}%
\label{CF302}}
\caption{Experiment setup in two indoor environments. (a) Meeting Room. (b) Classroom.}
\label{Wall_Tx_CD634_CF302}
\end{figure}

To expand the sensing coverage area inside the room and mitigate the negative effects of interference outside the room, based on the simulation model, we set the distance between the wall and the transmitter within the range of 1 m to 1.5 m. This deployment strategy is compared with a benchmark scenario where the transceivers are positioned at the center of the room. For each environment, in the benchmark setting, we maintain a larger distance of over 2.5 m between the wall and the transmitter. This comparison is motivated by the observation that, according to the proposed model, the influence of walls becomes negligible when the wall-device distance exceeds a certain threshold.  

In this case, we conduct experiments in two distinct environments with different deployments, as depicted in Fig. \ref{Wall_Tx_CD634_CF302}. The green-colored topologies that are deployed at the center of the venues serve as benchmarks. The orange-colored topologies represent the four testing scenarios, which should generate larger sensing regions compared to the benchmarks. We removed unnecessary furniture in these two environments to reduce the impact of unrelated factors on static power.

\begin{figure}[h]
  \centering
  \includegraphics[width=0.9\linewidth]{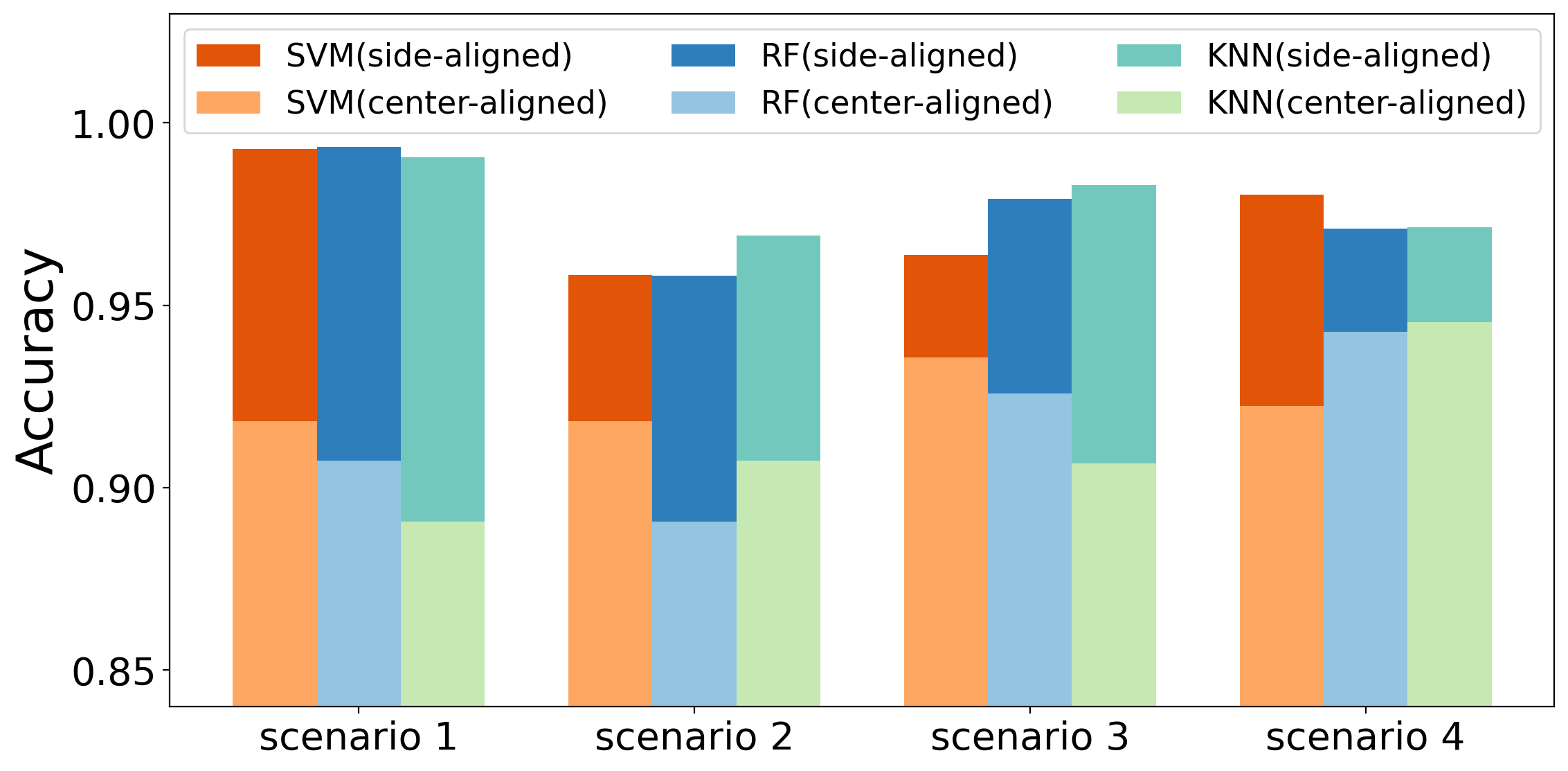}
  \caption{Comparison of the counting accuracy under single-wall-proximity topologies and benchmarks in various scenarios.}
  \label{Wall_Tx_various_scenarios}
\end{figure}

Fig. \ref{Wall_Tx_various_scenarios} presents a comparison of the accuracy results for these scenarios. Each scenario includes a comparison between the benchmark configurations (center-aligned) and the topologies leveraging wall-reflected signals (side-aligned) using the three machine learning models (SVM, RF, KNN). The light-colored bars represent the benchmarks, while the dark-colored bars indicate the improved accuracy attributed to the wall effect. Across all scenarios, we observe that the accuracy of cases where the transmitter is positioned closer to the wall (both light- and dark-colored) is consistently higher than that of the benchmark case (light-colored), with an increase from 2.1\% to 11.2\%. This result also serves as further confirmation of the property that the wall has a positive effect on the sensing coverage when the distance between the wall and the device is within a certain range.

{\bfseries Case 2: Scenarios where both devices are positioned close to the wall}

\begin{figure}[h]
\centering
\subfloat[]{\includegraphics[width=1.7in]{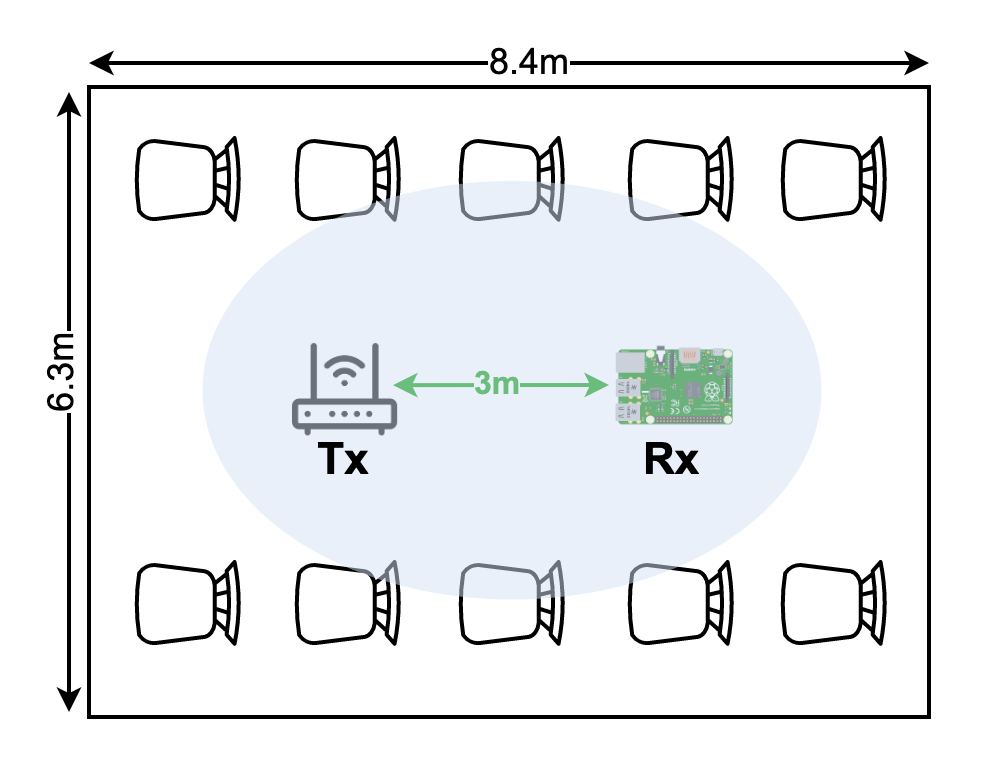}%
\label{CD634_3m}}
\subfloat[]{\includegraphics[width=1.7in]{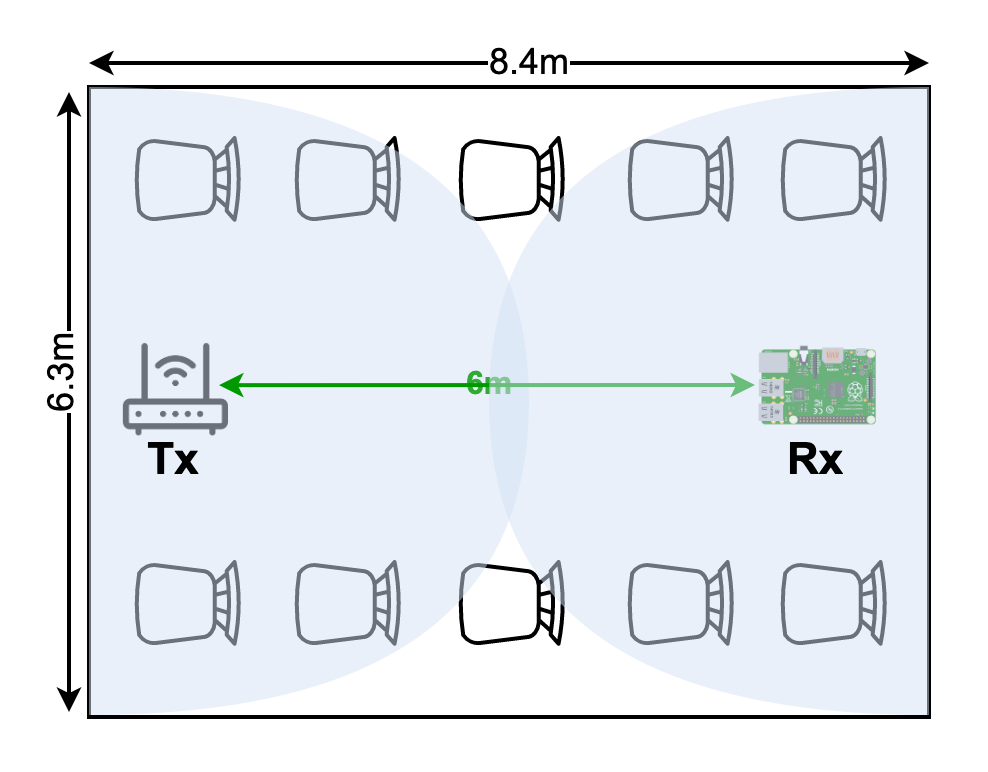}%
\label{CD634_6m}}
\caption{Enlarging sensing coverage with devices in two sides (Meeting Room). (a) Benchmark. (b) Wall-proximity topology.}
\label{Boosting_CD634_horizontal}
\end{figure}

\begin{figure}[h]
\centering
\subfloat[]{\includegraphics[width=1.7in]{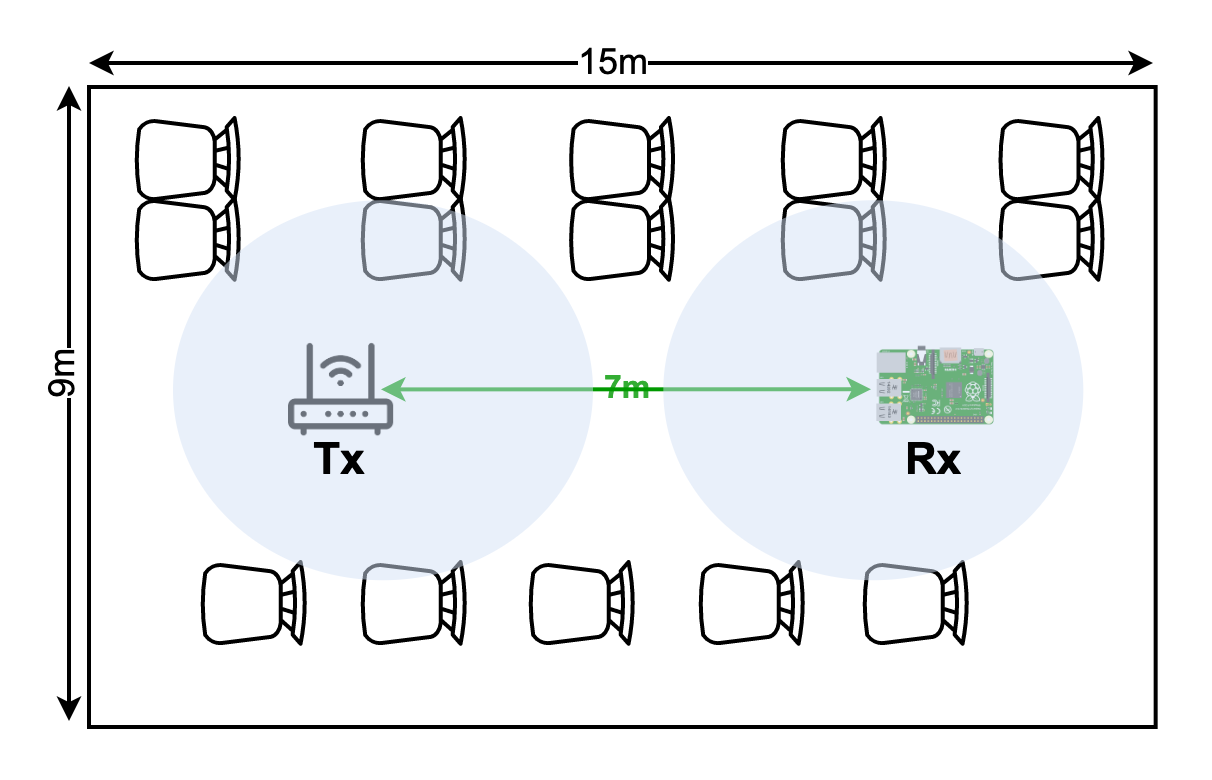}%
\label{AG206_7m}}
\subfloat[]{\includegraphics[width=1.7in]{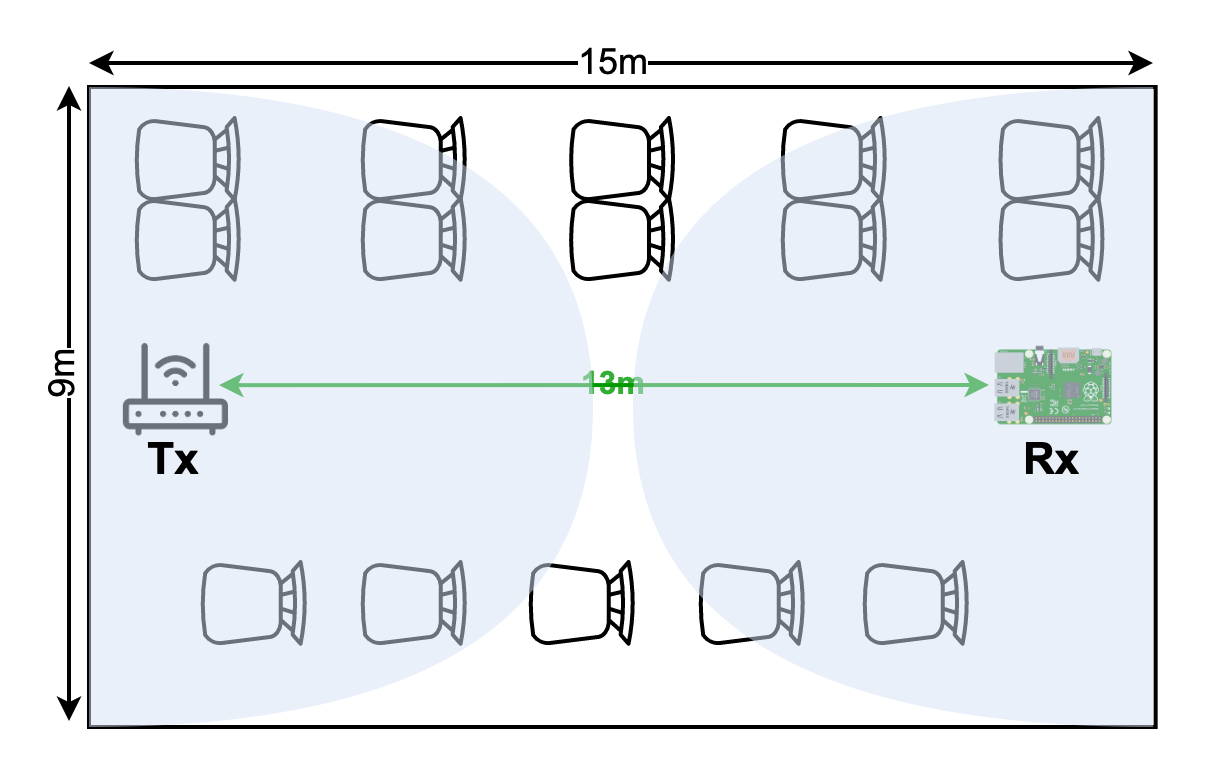}%
\label{AG206_13m}}
\caption{Enlarging sensing coverage with devices in two sides (Classroom). (a) Benchmark. (b) Wall-proximity topology.}
\label{Boosting_AG206}
\end{figure}

\begin{figure}[h]
\centering
\subfloat[]{\includegraphics[width=1.7in]{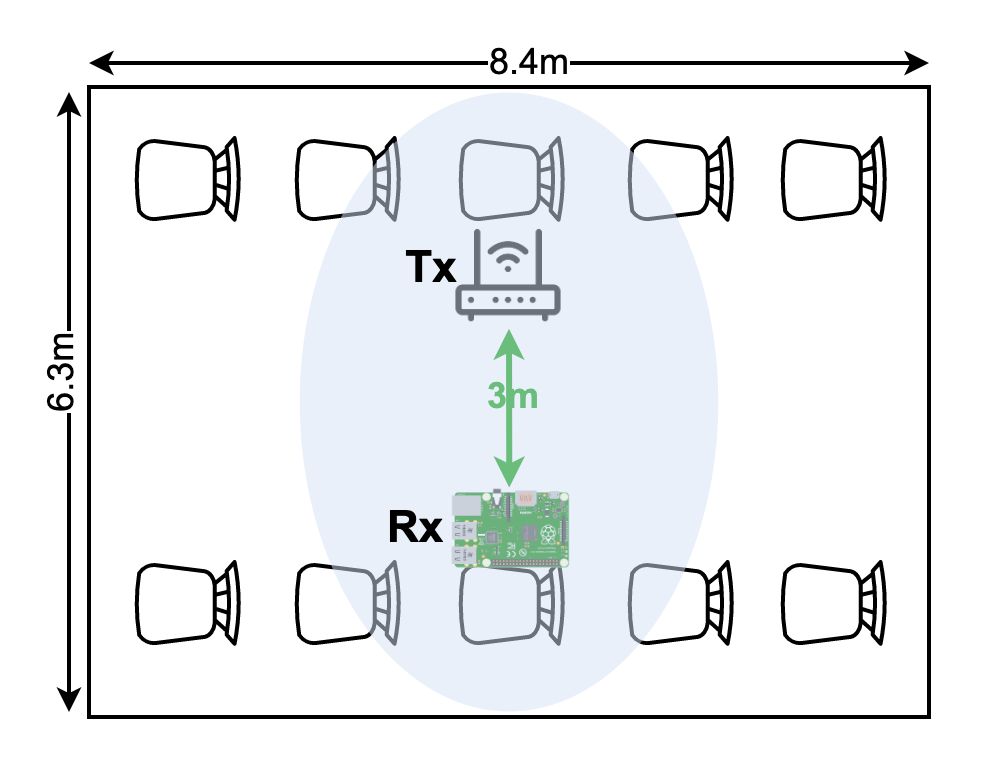}%
\label{CD634_center}}
\subfloat[]{\includegraphics[width=1.7in]{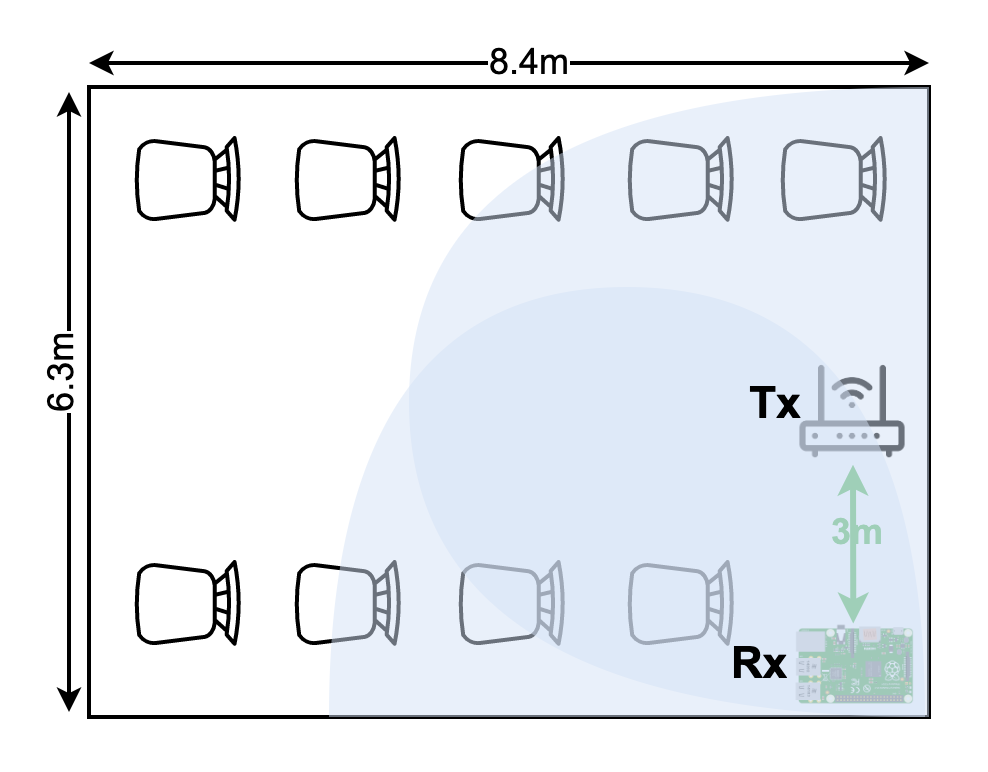}%
\label{CD634_nearwall}}
\caption{Enlarging sensing coverage with devices in the same sides. (a) Benchmark. (b) Wall-proximity topology.}
\label{Boosting_CD634_vertical}
\end{figure}

It is evident that the sensing area of the Wi-Fi system is quite limited if the placement of devices is not carefully planned. As shown in Fig. \ref{CD634_3m}, even when we deployed the optimal distance of 3 m between the transceivers in the meeting room, the sensing coverage did not encompass the entire room. To further enhance the sensing coverage, deploying both devices in close proximity to the walls can be a viable strategy, as illustrated in Fig. \ref{CD634_6m}. To evaluate the effect of this deployment strategy, we compare the sensing coverage by evaluating the accuracy of stationary crowd counting under three different scenarios:

\begin{enumerate}
\item In the first scenario, as shown in Fig. \ref{Boosting_CD634_horizontal}, the transceivers are horizontally placed in a meeting room. To expand the sensing coverage, we move the Tx and Rx in opposite directions within the room and deploy them near two sides of the room (walls).
\item In the second scenario, we maintain the same settings as in the first scenario but in a different environment, a classroom with a different size compared to the meeting room. As a result, the distance between the transceivers is distinct from the first scenario. This scenario is illustrated in Fig. \ref{Boosting_AG206}.
\item In the third scenario, as depicted in Fig. \ref{Boosting_CD634_vertical}, the transceivers are positioned vertically. In this case, we employ a different strategy to expand the sensing coverage. The Tx and Rx are moved synchronously, maintaining a fixed distance between them, and positioned close to the same side of the room (wall).  
\end{enumerate}

In each of the mentioned scenarios, we compare the deployment strategy with a benchmark where the transceivers are placed at the center of the room. This benchmark serves as a reference point for evaluating the effectiveness of boosting the sensing coverage through different deployment strategies.

\begin{figure}[h]
  \centering
  \includegraphics[width=0.9\linewidth]{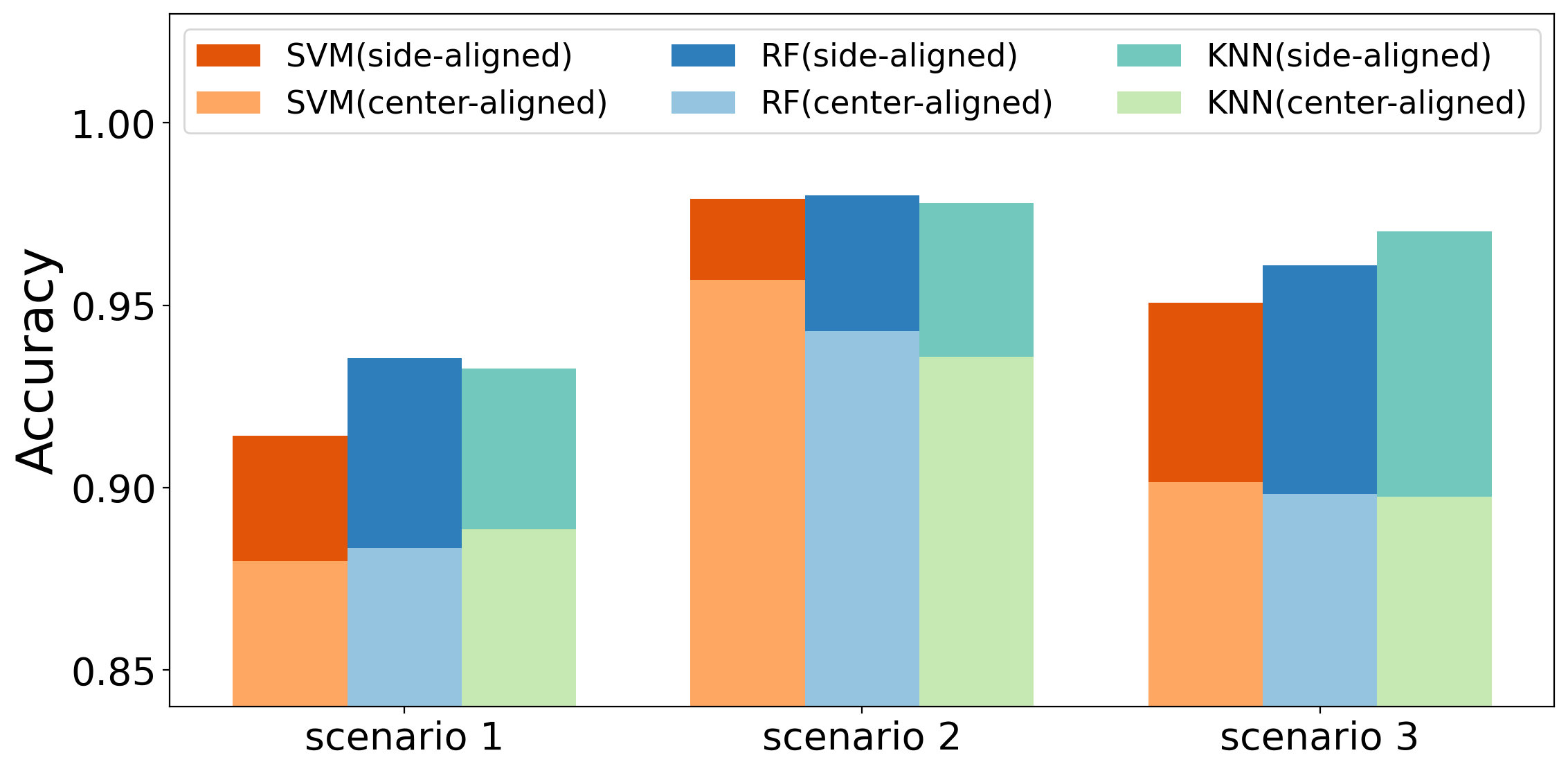}
  \caption{Comparison of the counting accuracy under double-wall-proximity topologies and benchmarks in various scenarios.}
  \label{boosting_rf_various_scenario}
\end{figure}

Based on the results in Fig. \ref{boosting_rf_various_scenario}, deploying devices near the room boundaries (dark-colored) significantly enhanced crowd counting accuracy. Even when the distance between the transceivers exceeds the optimal range relative to the benchmark, accuracy improvements of 2.1\% to 7.8\% were observed across all three scenarios. This outcome underscores the effectiveness of our deployment strategies in enhancing both sensing coverage and accuracy.

Overall, the experimental observations in Sections \ref{subsection_RM} and \ref{cc_1} exhibit strong consistency with the simulation-based analyses presented in Section \ref{Section_analysis}. As illustrated in Figs. \ref{fig:RR_SSNR_compare}, \ref{fig:Wall_Tx_compare}, and \ref{Tx_Rx_various_distances}, the simulated effects of wall–Tx distance and Tx–Rx distance on SSNR and sensing area align well with the observed trends in the corresponding real-world performance metrics, including MAE of respiratory rate and crowd counting accuracy. These results collectively validate the theoretical models and simulation assumptions adopted in this work. Furthermore, the extended deployment scenarios evaluated in Section \ref{cc_2} demonstrate that these modeled properties can be effectively applied in more complex real-world settings, further supporting their practical value in boosting sensing coverage.

\section{Discussion}

In this section, we acknowledge the limitations of our work and highlight potential future directions for further improvement and exploration.

{\bfseries Multiple environments.}
Although the experiments were conducted in three representative indoor environments (a meeting room and two classrooms) with five participants, these setups were carefully designed to ensure controlled and repeatable measurements across different transceiver placements. The proposed framework is general in formulation and can be extended to other room geometries and materials; however, further validation in more diverse environments (e.g., furniture-rich or metallic spaces) will be pursued in future work to strengthen the generalization of the findings.

{\bfseries Multiple sensors.}
The current work focuses on a single pair of transceivers, which may be inadequate for large spaces or multiple targets. To address this, deploying multiple sensors with suitable topologies based on our model \cite{multisensor2018,multisensor2023} is recommended. By optimizing placement near walls and enhancing coverage through strategic positioning, a network of sensors can be effectively distributed to improve sensing capabilities. Future research should explore optimal topologies and deployment strategies for multiple sensors, ensuring comprehensive sensing coverage in scenarios where a single pair is insufficient. 

{\bfseries Other factors for optimal device deployment.}
In this study on stationary crowd counting, targets were randomly positioned throughout the space, without explicitly considering the impact of their locations on sensing capabilities. It is important to note that in real-world scenarios, optimal device performance may not always involve placing devices near walls, as other factors, such as target positioning, obstacles, and device movement direction, can also influence sensing outcomes. Future work could involve conducting experiments in a wider range of environments, incorporating various room sizes, layouts, and wall materials, as well as directionally diverse device movements. By integrating these factors into deployment strategies, we can further generalize the proposed framework and deepen our understanding of how device placement affects sensing performance, thereby supporting more accurate and practical applications.

\section{Conclusion}

Our work highlights the significant impact of both the wall-device distance and transmitter-receiver distance on Wi-Fi sensing coverage in indoor environments. By introducing a theoretical model, we have successfully demonstrated the relationship between wall-device distances and sensing coverage. The performance of our proposed model was evaluated through experiments on respiration monitoring and stationary crowd counting across representative indoor settings. The insights gained from this model enable us to strategically plan device placement, ultimately expanding the sensing coverage in practical deployments of wall-reflection sensing systems. This research successfully bridges the gap between theoretical understanding and practical implementation, paving the way for deployment-aware Wi-Fi sensing in typical indoor scenarios. Our findings contribute to advancing the field of Wi-Fi sensing and offer valuable guidance for optimizing sensing performance in realistic indoor environments


\bibliographystyle{IEEEtran}

\end{document}